\documentclass[bibyear]{aa}
\usepackage{txfonts}
\usepackage{graphicx}
\usepackage[usenames,dvipsnames]{color}

\usepackage{hyperref}
\usepackage{natbib}
\usepackage{color}
\definecolor{cyan}{rgb}{0,0.8,0.8}
\definecolor{red}{rgb}{1,0,0}
\definecolor{green}{rgb}{0,1,0}
\definecolor{blue}{rgb}{0,0,1}
\definecolor{magenta}{rgb}{1,0,1}

\begin{document}

\title{Hot UV-bright stars of galactic globular clusters\thanks{Based on
    observations with the ESO Very Large Telescope at Paranal
    Observatory, Chile (proposal ID 089.C-0210)}\fnmsep\thanks{The extracted
    spectra, their best-fitting model spectra, and the evolutionary
    tracks used in this paper will be available at CDS.}}
\author{S.\,Moehler\inst{1}
\and W.\,B.\,Landsman\inst{2}
\and T.\,Lanz\inst{3}
\and M.\,M.\,Miller Bertolami\inst{4,5}
} 
\institute{European Southern Observatory,
Karl-Schwarzschild-Str. 2, D 85748 Garching, Germany, \email{smoehler@eso.org}
\and
Adnet Systems, NASA Goddard Space Flight Center, Greenbelt, MD 20771,
USA 
\and Observatoire de la C\^ote d'Azur, F-06304, Nice, France
\and Instituto de Astrof\'isica de La Plata, UNLP-CONICET, Paseo del Bosque s/n, 1900 La Plata, Argentina
\and Facultad de Ciencias Astron\'omicas y Geof\'isicas, Universidad Nacional de la Plata, Paseo del Bosque s/n, 1900 La Plata, Argentina
} 
\date{Received April 15, 2019 / Accepted May 7, 2019}

\abstract
{{We have performed a census of the UV-bright population in 78
    globular clusters using wide-field UV telescopes.  This population
    includes a variety of phases of post-horizontal branch (HB)
    evolution, including hot post-asymptotic giant branch (AGB) stars,
    and post-early AGB stars. There are indications that old stellar
    systems like globular clusters produce fewer post-(early) AGB
    stars than currently predicted by evolutionary models, but
    observations are still scarce.}}
{We wish to derive effective temperatures, surface gravities, and
  helium abundances of the luminous hot UV-bright stars in these
  clusters to
  determine their evolutionary status and compare the observed numbers
  to predictions from evolutionary theory.}
{We obtained FORS2 spectroscopy of eleven of these UV-selected objects
  (covering a range of $-2.3<$[Fe/H]$<-$1.0), which we (re-)analysed
  together with previously observed data. We used model
  atmospheres of different metallicities, including super-solar ones.
  Where possible, we verified our atmospheric parameters using UV
  spectrophotometry and searched for metal lines in the optical spectra.
  We calculated evolutionary sequences for four metallicity regimes and
  used them together with information about the HB morphology of the
  globular clusters to estimate the expected numbers of post-AGB
  stars.}
{ We find that metal-rich model spectra are required to analyse stars
  hotter than 40\,000\,K.  Seven of the eleven new luminous UV-bright
  stars are post-AGB or post-early AGB stars, while two are evolving
  away from the HB, one is a foreground white dwarf, and another is a white
  dwarf merger. Taking into account published information on other hot
  UV-bright stars in globular clusters, we find that the number
  of observed hot post-AGB stars generally agrees with the predicted
  values, although the numbers are still low. }
{Spectroscopy is clearly required to identify the evolutionary status of hot
UV-bright stars. For hotter stars, metal-rich model spectra are required
to reproduce their optical and UV spectra, which may affect the
flux contribution of hot post-AGB stars to the UV spectra of evolved
populations. While the observed numbers of post-AGB and post-early AGB
stars roughly agree with the predictions, our
    current comparison is affected by low number statistics. }
{}
\maketitle
\section{Introduction}
\label{sec:intro}
Ultraviolet images of globular clusters are often dominated by one or
two hot, luminous, UV-bright stars that are more than a magnitude brighter than
the horizontal branch ({\bf HB}) and hotter than 7000\,K. Such stars
  are in various evolutionary stages after helium core burning ({\bf
    HeCB}). These evolutionary stages are the most uncertain phases
of the evolution of low-mass single stars, caused by uncertainties in
the size of the C/O-core due to our poor understanding of convective
boundary mixing \citep{Charpinet+11, Constantino+15}, and our
lack of understanding of winds in red giant branch ({\bf RGB}) stars
\citep{McDonald+15,Salaris+16}.  The previous mass loss on the  RGB
plays an important role here because it 
determines whether a star even ascends the asymptotic giant branch
  ({\bf AGB}): HB stars with very low hydrogen
envelope masses evolve directly from the extreme horizontal
  branch ({\bf EHB}, T$_{\rm eff}\gtrsim$ 20\,000\,K) to the white dwarf
stage, whereas stars with hydrogen envelope masses of more than
0.02\,M$_\odot$ will at least partly ascend the AGB.  Owing to the
previous uncertainties, both the masses of the C/O-core and the H-rich
envelope of post-HeCB are loosely constrained. In particular,
theoretical post-AGB tracks have only been very little tested for old
low-mass stars because suitable observations are scarce. A similar
situation holds for post-EHB stars that evolve directly to the white
dwarf phase and avoid the AGB. As an example, \citet{Brown+08} found
far fewer hot post-AGB stars than expected in their Space Telescope Imaging Spectrograph (STIS) UV imagery of
M\,32, while \citet{Weston+10} reported a similar lack of hot
post-AGB stars in the Galactic halo. The status of hot post-AGB stars
in globular clusters remains uncertain: with only about one hot
post-AGB star per cluster, many clusters must be observed in order to
compare number counts, luminosities, and surface gravities with evolutionary
tracks.

The detection of hot post-AGB stars in optical colour-magnitude
diagrams is limited by selection effects that are caused by crowding in the
cluster cores and by the large bolometric corrections for these hot
stars. More complete searches are possible for hot post-AGB stars in
planetary nebulae, for example, by using O\,{\sc iii} imaging, as
performed by \citet{Jacoby+97} and \citet{Bond15}. The
  meaningfulness of such searches, however, is limited by the long
  evolutionary timescales of low-mass post-AGB stars, which make the
  appearance of a planetary nebula unlikely.  Only four
  planetary nebulae have indeed thus far been discovered in the Galactic
  globular cluster system \citep{Jacoby+97}.

\citet{Renzini85} and \citet{deBoer87} first pointed out the advantages of
wide-field UV imagery to obtain a complete sample of hot
post-AGB stars in a large number of globular clusters.  The
UV suppresses the dominant cool-star population and
emphasizes the hot stars, ensuring that all hot post-AGB stars in the
cluster are detected.  

Fourteen globular clusters were observed with Ultraviolet
  Imaging Telescope ({\bf UIT}, \citealt{Stecher+97}) in 1990 and 1995 at
  1620\,\AA\ with a $40'$ field of view ({\bf FOV}).  \citet{Moehler+98} used
  ground-based spectroscopy to derive temperatures and gravities of
  the newly discovered hot UV-bright stars. Their results clearly
  illustrated the points made above, namely that the HB morphology has
  a strong influence on the evolutionary status of hot UV-bright stars
  and that optical selections suffer from a strong bias towards the
  most luminous hot stars.  Subsequently, the {Galaxy Evolution
    Explorer 
({\bf GALEX})} satellite
  \citep{Martin+05} was used to obtain wide-field UV images
  of 41 globular clusters at 1520\,\AA\ with a $1.2^\circ$ diameter FOV.
  Because {GALEX} observations are limited by its brightness
  constraints to mostly high Galactic latitude fields, we obtained
  archival UV images of an additional 31 clusters using the
  Swift Ultraviolet-Optical Telescope ({\bf UVOT}, \citealt{Poole+08}).  The
  UVOT images are less satisfactory because its FOV is smaller ($17' x 17'$)
and its wavelength is longer.  The solar-blind UVM2 filter on the UVOT has
  a central wavelength of 2250\,\AA,\, while the UVW2 filter has a
  central wavelength of 1930\,\AA.  The UV imagery from all
  three wide-field telescopes was complemented by UV imagery
  with the Hubble Space Telescope ({\bf HST}, \citealt{Nardiello+18}), which
  provides much higher spatial resolution and photometric precision,
  but rarely encompasses the entire globular cluster within its FOV.

  This brings the total number of globular clusters with UV imaging
  data to 78. Among these, we recovered all previously known hot
  post-AGB star candidates and found 19 new hot post-AGB star
  candidates and 16 other new hot stars (mainly apparent EHB or
  post-EHB stars in globular clusters where only red HB stars have
  been known so far). Because each cluster contains only a few post-AGB stars, a large sample of clusters is needed to test stellar
  evolution predictions. To this aim, we started a project to study the
  whole sample of globular clusters with UV imagery and compare it
  with stellar evolution predictions.  We are currently working on a
  paper describing the results of this greatly enlarged sample and the
  new hot post-AGB candidates. As a first step, we describe here the
  optical spectroscopic observations of 10 of the brightest hot
  post-AGB star candidates in the Southern Hemisphere and one post-EHB
  star candidate in M\,4. The paper is organised as follows. First we describe in
  Sect.\,\ref{sec:obs_proc} the new observations as well as
 unpublished data from \cite{Moehler+98}, and in
  Sects.\,\ref{sec:rv} and \ref{sec:param} we derive stellar parameters
  for all the stars in our sample and discuss some particular
  cases. This sample is enlarged in Section \ref{sec:lit} by including
  all UV-bright stars that have been studied in previous publications. In
  Sect.\,\ref{sec:masses} we derive masses from the atmospheric
  parameters and the distances of the globular cluster, and we discuss the
  effect of {\it Gaia} results for some of the closest clusters.  In Sect.\,\ref{sec:evol} we then present stellar evolution models of
  post-HeCB stars at different metallicities and for different masses
  on the HB and compare simple estimates of the number
  of luminous hot UV-bright stars predicted by the models with those
  observed in our sample. Finally, we end the paper in Sect.\,\ref{sec:conclusions}  with some preliminary conclusions and discuss what
  needs to be done to improve the comparison.


\begin{table*}[!h]
\caption[]{Target coordinates and brightness.\label{tab:target}}
\begin{tabular}{llllrclc}
\hline
\hline
globular & star & $\alpha_{2000}$ & $\delta_{2000}$ & $m_{UV}$ & UV Tel. & $V$ & Ref \\
cluster  &      &  [h:m:s]        & [$^\circ$:\arcmin:\arcsec]  & [mag]  & &[mag] & \\
\hline
NGC\,5139     & UIT151  & 13:27:05.06 & $-$47:21:56.6 & 15.29\tablefootmark{1}& UIT & 16.577& (1) \\
              & UIT644  & 13:26:44.96 & $-$47:27:09.5 & 13.12\tablefootmark{1}& UIT & 12.713 & (1)  \\
              & UIT1275 & 13:26:52.77 & $-$47:29:44.1 & 14.75\tablefootmark{1}& UIT & 14.306 & (1)  \\
              & UIT1425 & 13:26:11.56 & $-$47:30:49.2 & 15.31\tablefootmark{1}& UIT & 15.660 & (1)   \\
              & ROA5342 & 13:25:45.42 & $-$47:24:02.0 & 14.73\tablefootmark{1}& UIT & 15.944 & (1) \\
              & Dk3873  & 13:26:13.95 & $-$47:25:30.1 & 15.29\tablefootmark{1}& UIT & 16.531 & (1)  \\
NGC\,6121     & UVBS2   & 16:23:26.25 & $-$26:31:27.0 & 15.07\tablefootmark{1} & UIT & 16.388& (2) \\
NGC\,6656     & UVBS2   & 18:36:22.85 & $-$23:55:19.3 & 14.88\tablefootmark{2} & Swift UVOT & 14.533 & (3) \\
NGC\,6712     & C49     & 18:53:03.45 & $-$08:42:30.5 & 17.00\tablefootmark{2} & Swift UVOT & 16.72 & (4)\\
NGC\,6779     & ZNG2    & 19:16:41.28 & $+$30 12 48.5 & 14.40\tablefootmark{2} & Swift UVOT & 15.13 & (5) \\
NGC\,7099     & UVBS2   & 21:40:18.08 & $-$23:13:22.3 & 15.54\tablefootmark{3} & GALEX & 16.75 & (6) \\
\hline
\multicolumn{8}{c}{Unpublished data from \cite{Moehler+98}}\\
\hline
NGC\,5139 & ROA542  & 13:25:49.91 & $-$47:22:59.7 & 15.33$^1$ & UIT & 12.876 & (1) \\
          & ROA3596 & 13:27:46.09 & $-$47:30:57.6 & 14.73$^1$ & UIT & 14.179 & (1) \\
\hline
\end{tabular}
\tablefoot{
\tablefoottext{1}{B5 (AB mag, $\lambda_c$=1615\,\AA, FWHM=225\,\AA)}
\tablefoottext{2}{W2 (AB mag, $\lambda_c$=1928\,\AA, FWHM=657\,\AA)}
\tablefoottext{3}{NUV\_AB (AB mag, $\lambda_c$=1528\,\AA, FWHM=228\,\AA)}
}
\tablebib{(1) \citet{Bellini09}; (2) \citet{Mochejska+02}; (3) \citet{Monaco+04}; (4) \citet{Cudworth88}; 
(5) \citet{Rosenberg00}; (6) \citet{Sandquist99}} 
\end{table*}

\section{Observations and data reduction}\label{sec:obs_proc}
The coordinates and brightness of our targets are listed in
Table\,\ref{tab:target}. The data were taken with the FOcal Reducer
and low-dispersion Spectrograph 2 (FORS2, \citealt{Appenzeller+98}) at
the ESO Very Large Telescope (VLT) UT1. We used the grism GRIS\_1200B with a slit width of
0\farcs5 and verified the resolution from the arc lamp
  images. The stars in $\omega$\,Cen (except for $\omega$\,Cen$-$UIT151)
were observed in multi-object spectroscopy (MOS) mode with movable
slitlets of about 20\arcsec\ length and have a spectral
  resolution of about 1.7\,\AA. All other stars (including
$\omega$\,Cen$-$UIT151) were observed with the long slit with a length
of 6\farcm8 and have a spectral resolution of about
  1.4\,\AA. We always took two exposures of each target for
consistency checks.

The data were observed between April 1 and July 22, 2012 (date at the
beginning of the night). They were processed soon afterwards with the
\href{http://www.eso.org/sci/software/pipelines/}{ESO FORS2 pipeline}
(version fors-4.12.8), except for the flux calibration (see
  Appendix\,\ref{app:flux}). For the bias correction
the bias applied to science and calibration data was scaled with the
prescan level of these data. In order to take possible structure in
the bias into account, the full (prescan-corrected) master bias was
subtracted and not just a number. The flat fields were summed per
setup and date and normalized using a smoothing with a radius of 10 pixels
along both axes. The wavelength calibration used 15 arc lines for the
long-slit spectra and 15--17 arc lines for the MOS spectra (depending
on the slitlet location). The average accuracy achieved in this way was
0.1\,pixel for the long-slit spectra and 0.24\,pixel for the MOS
data. We used local sky subtraction, that is, fitting of the sky spectrum
per slit on the two-dimensional spectra (before rectification and
rebinning). Because of the brightness of our targets and the blue
wavelength range, the sky level was rather low, even though most of the
data were observed while the Moon was above the horizon. To verify the
quality of the sky subtraction, we checked our spectra for the presence
of G-band features, which should not be present in these hot stars, and
found none. The details of the response determination for this
volume-phase holographic grism are described in
Appendix\,\ref{app:flux}.

\begin{figure}
\includegraphics[height=\columnwidth,angle=270]{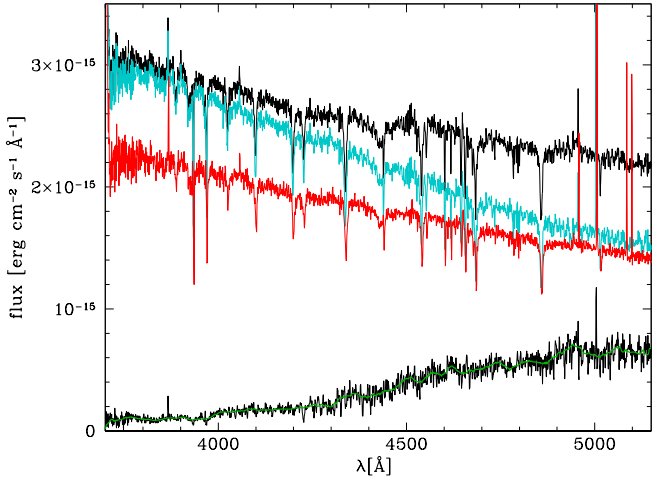}
\caption[]{Flux-calibrated spectra of NGC\,6656$-$UVBS2 and its
  neighbour (top and bottom black spectra), the smoothed spectrum of
  the neighbour (green), and the corrected spectrum of
  NGC\,6656$-$UVBS2 (cyan). The spectrum extracted with narrow
    limits is marked in red.}\label{fig:M22$-$UVBS1}
\end{figure}

\begin{figure*}
\includegraphics[width=\textwidth]{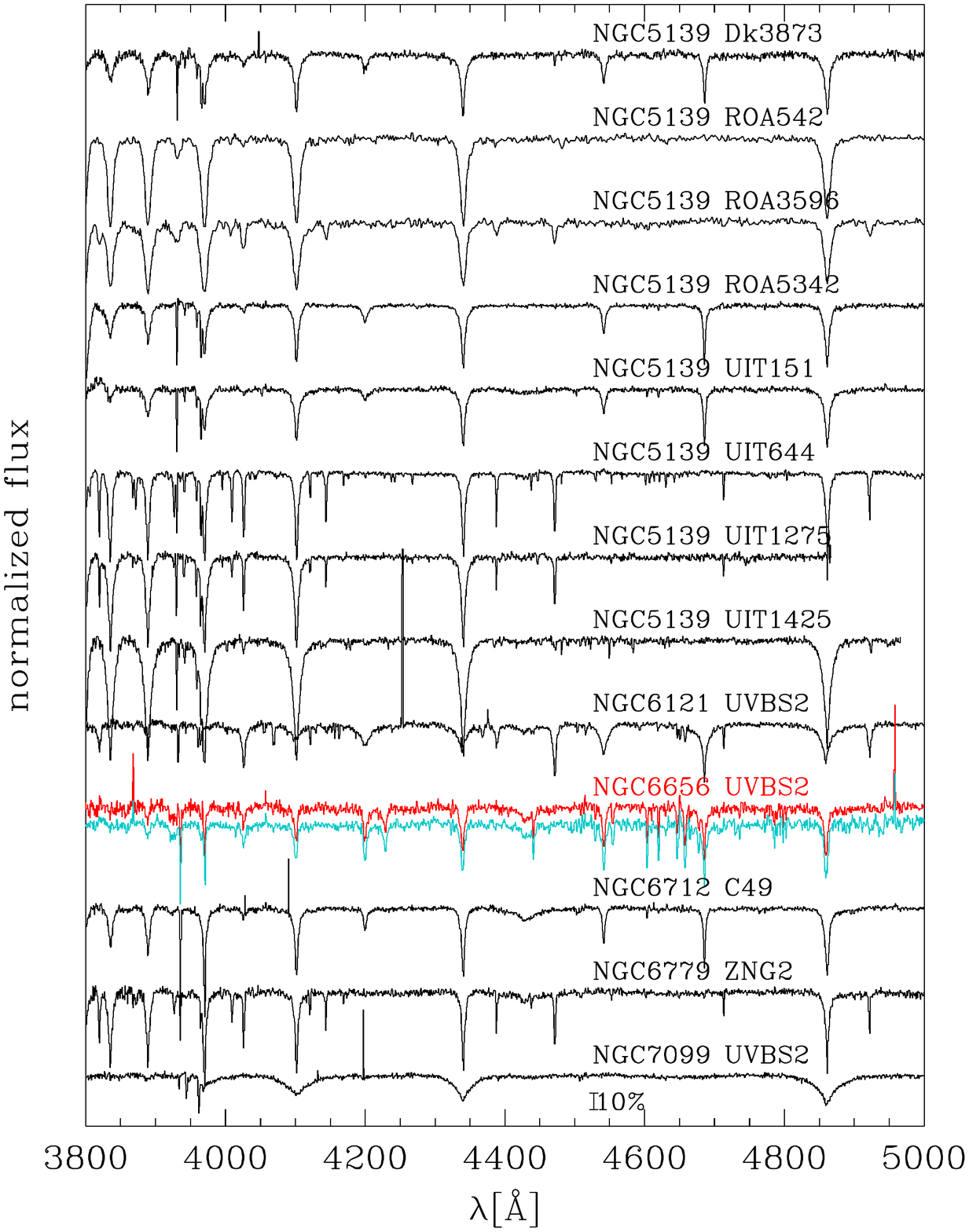}
\caption[]{Normalized spectra. NGC\,5139$-$ROA542 and
    NGC\,5139$-$ROA3596 are the unpublished spectra from
    \citet{Moehler+98}.
For NGC\,6656$-$UVBS2 we show the
  results of both extraction methods (red marks the extraction within
  a narrow window, and cyan marks the corrected
    spectrum, offset by 0.1 along the y-axis). The emission lines for that
  spectrum are real and due to its planetary nebula.
}\label{fig:spec}
\end{figure*}

The extracted spectra showed that NGC\,6656$-$UVBS2 has
a close cool fainter neighbour, whose spectrum overlaps with that of
the intended target. We did not see any significant lines in the
neighbour's spectrum, therefore we smoothed it with a median filter of 14\,\AA\,
half-width and subtracted it from the spectrum of NGC\,6656$-$UVBS2
(see Fig.\,\ref{fig:M22$-$UVBS1}). In another attempt to avoid the
contamination from the cool neighbour we extracted only
the part of the spectrum away from the neighbour (marked in red in
Fig.\,\ref{fig:spec}). The different slopes of the two spectra
  point towards possible over- and/or undercorrection of the flux from
  the cool neighbour. The ratio of the two spectra shows no noticeable
line residuals. Considering the experience with crowded field
spectroscopy recorded in \cite{mosw06} , we prefer to use the narrowly
extracted spectrum for further analysis.

\subsection{Unpublished data from \cite{Moehler+98}}
After discovering two unpublished spectra (NGC\,5139$-$ROA542 and
NGC\,5139$-$ROA3596) from the same observing run as \cite{Moehler+98}, we
decided to analyse all the data from that observing run again. Based on
comments from V. Dixon and P. Chayer, we checked the resolution of
these data and found that the line width used in \cite{Moehler+98} was too
high. A pixel scale of 0.336\arcsec/pixel and dispersion of
100\,\AA/mm yield a resolution of 5.6\,\AA\ and not 6.7\,\AA,\ as used
by \cite{Moehler+98}. The lower value explains most of the
differences between the results reported here and those from
\cite{Moehler+98}.

\section{Radial velocities}\label{sec:rv}

To correct the observed spectra to laboratory wavelengths, we first
corrected to the heliocentric system using the \,{\sc midas} command
\,{\sc compute/barycor}. Then we determined the radial velocities by
fitting Gaussian profiles to the cores of strong lines using the \,{\sc
  midas} command \,{\sc center/gauss}. We selected the lines and their
fit range manually to ensure that no surviving cosmics or noise peaks
distort the results. We did not use lines at wavelengths below
3800\,\AA\ as the bluest arc line in our calibration data is at
3888\,\AA. The resulting radial velocities are listed in
Table\,\ref{tab:params}, together with the rms error of the individual
measurements and the velocities of the
corresponding clusters from the literature. The one glaring
discrepancy is the radial velocity for NGC\,7099$-$UVBS2 whose value
suggests that this star is not a cluster member. This is also
supported by the parameters derived in Sect.\,\ref{sec:param}, which
argue in favour of a foreground white dwarf.

Two stars with very small formal errors ($\lesssim$10\,km\,s$^{-1}$)
show velocity differences well above these errors: NGC\,6656$-$UVBS2
($+$33.6\,km\,s$^{-1}$) and NGC\,5139$-$UIT1275
($+$40.5\,km\,s$^{-1}$).  We recall, however, that one
pixel in our data corresponds to 48\,km\,s$^{-1}$ and the resolution
of our data corresponds to 90--120\,km\,s$^{-1}$. Therefore we decided to
consider all stars except for NGC\,7099$-$UVBS2 as members of their
respective globular clusters.
\begin{table*}
\caption[]{Atmospheric parameters and heliocentric velocities of the
  stars. The status acronyms are p-AGB (post-AGB), post-early AGB
  (peAGB), post-HB (pHB), white dwarf merger (WDM), and white dwarf
  (WD). We have no velocity information for the data from
  \cite{Moehler+98}}\label{tab:params} 
\begin{tabular}{llrrrr|c|rrrl}
\hline
\hline
cluster & star & $T_{\rm eff}$ & $\log g$ & $\log {\frac{n_{\rm He}}{n_{\rm H}}}$ & $M$
& status & $v_{\rm hel}$ & $n_v$ & $v_{\rm cluster}$ & Ref.\\
 & & [K] & [cm s$^{-1}$] & & [M$_\odot$] & & [km\,s$^{-1}$] & & [km\,s$^{-1}$] &\\
\hline
NGC\,5139 & UIT151  & 62600$\pm$2400   & 5.11$\pm$0.10& $-$0.99$\pm$0.08& 0.24 & peAGB & $+$231.1$\pm$12 &  9 &$+$231.8 & (1)\\
     & UIT644  & 18000$\pm$\ \ 800& 2.82$\pm$0.10& $-$0.35$\pm$0.06& 0.29 & pAGB & $+$232.3$\pm$13   & 26 & & \\
     & UIT1275 & 17200$\pm$1100& 3.34$\pm$0.18& $-$1.27$\pm$0.18& 0.25 & pHB & $+$272.3$\pm$\ \ 6& 14&  & \\
     & UIT1425 & 11700$\pm$\ \ 200& 3.57$\pm$0.08& $-$1.72$\pm$0.36& 0.23 & pHB & $+$213.6$\pm$17   &  9&  & \\
     & Dk3873  & 49700$\pm$1800& 4.99$\pm$0.12& $-$1.21$\pm$0.12& 0.28 & peAGB & $+$231.2$\pm$26   & 11&  & \\
& ROA5342 & 52000$\pm$2000   & 4.79$\pm$0.10& $-$1.17$\pm$0.10& 0.29 & peAGB & $+$227.6$\pm$10   & 10&  & \\[1mm]
NGC\,6121 & UVBS2\tablefootmark{1} & 46600$\pm$\ \ 600& 5.82$\pm$0.12& $+$0.89$\pm$0.18&0.85& WDM & $+$74.6$\pm$15   & 17& $+$71.5 & (2)\\
NGC\,6656 & UVBS2   & 78800$\pm$9900   & 4.58$\pm$0.20& $+$0.36$\pm$0.16&0.18& pAGB & $-$111.3$\pm$\ \ 4&  3& $-$144.9 & (2)\\
NGC\,6712 & C49\tablefootmark{2} & 53300$\pm$4000   & 4.50$\pm$0.12& $-$1.04$\pm$0.12&0.31& pAGB & $-$108.0$\pm$\ \ 8&  8& $-$109.0 & (3)\\
NGC\,6779 & ZNG2    & 21500$\pm$1800   & 3.07$\pm$0.16& $-$0.85$\pm$0.08&0.19& pAGB & $-$123.5$\pm$\ \ 9& 15& $-$138.1 & (4) \\
NGC\,7099 & UVBS2   & 65000$\pm$2000   & 7.22$\pm$0.12& $-$3.29$\pm$0.12& & WD &  $+$11.4$\pm$\ \ 8&  3& $-$184.4 & (2)\\
\hline
\multicolumn{9}{c}{Unpublished data from \cite{Moehler+98}}\\
\hline
NGC\,5139 & ROA542  & 10600$\pm$\ \ 200 & 2.55$\pm$0.10 & $-$1.00 &0.34& pHB & & & &\\ 
          & ROA3596 & 17800$\pm$1000 & 3.56$\pm$0.16 & $-$1.26$\pm$0.18&0.41& pHB & & & & \\
\hline
\multicolumn{9}{c}{New analysis of data from \cite{Moehler+98}}\\
\hline
NGC\,2808 & C2946   & 24900$\pm$1800 & 4.78$\pm$0.20 & $-$1.88$\pm$0.14 &0.68& pHB & & & & \\
          & C2947   & 15000$\pm$1200 & 3.92$\pm$0.24 & $-$1.45$\pm$0.44 &0.41& pHB & & & & \\
          & C4594   & 19900$\pm$1600 & 3.79$\pm$0.22 & $-$1.63$\pm$0.26 &0.34& pHB & & & & \\
NGC\,6121 & Y453    & 54900$\pm$2000 & 5.62$\pm$0.14 & $-$1.25$\pm$0.10 &0.59& peAGB & & & & \\
NGC\,6723 & III60   & 43000$\pm$1400 & 4.72$\pm$0.14 & $-$1.19$\pm$0.14 &1.02& peAGB & & & & \\
          & IV9     & 25900$\pm$2000 & 4.02$\pm$0.22 & $-$1.06$\pm$0.10 &0.88& peAGB & & & & \\
NGC\,6752 & B2004   & 34500$\pm$\ \ 800 & 5.18$\pm$0.14 & $-$2.45$\pm$0.24 &0.35& pHB & & & & \\
\hline
\end{tabular}
\tablefoot{
\tablefoottext{1}{\citet{Mochejska+02} refer to the star as B2 and describe its spectrum as helium rich.}
\tablefoottext{2}{\citet{Remillard+80} classified the star as sdO.}
}
\tablebib{(1)~\citet{Johnson+08}; (2) \citet{Lane+10}; (3)
  \citet{Yong+08}; (4) \citet{Webbink81}.}
\end{table*}

\section{Atmospheric parameters}\label{sec:param}
Following \citet{Moehler+11} and \citet{Brown+12}, we analysed
the spectra with grids of non-local thermal equilibrium (NLTE) line-blanketed model atmospheres and
NLTE model spectra that were calculated with the NLTE model atmosphere
code TLUSTY \citep{HubenyLanz95} and the companion spectrum synthesis
code SYNSPEC \citep{HubenyLanz17}. The model atmosphere grids were tailored
to the studied sample with the parameter space delimited in
Table\,\ref{tab:models}. Over 1100 new model atmospheres have been
produced with steps in the model grids of 2000\,K in effective
temperature, 0.25\,dex in surface gravity, and 0.5\,dex in helium-to-hydrogen
abundance ratio by number, so that the observed spectra can be matched
with model spectra that were interpolated from the grids.  Scaled-solar
abundances based on \citet{Grevesse+98}
at the cluster's mean metallicities have been assumed, and
kept the same for models with various He/H abundance ratios. The
implied change in the total mass fraction of the heavy elements has
very limited consequences on our analysis (see \citealt{Moehler+11}).

The model atmospheres at higher temperature
($T_{\rm eff} \geq 40\,000$\,K) allow for departures from LTE for 1132
explicit levels and superlevels of 52 ions (H, He, C, N, O, Ne, Mg,
Al, Si, P, S, and Fe), as in \cite{Moehler+11}. A detailed description
of the model atoms and the source of the atomic data can be found in
\citet{LanzHubeny03,LanzHubeny07}. Microturbulent velocity is assumed
to be $v_{\rm t} = 5\,{\rm km\,s}^{-1}$.

At lower temperatures ($T_{\rm eff} \leq 30\,000$\,K), we naturally
considered lower ionization stages and excluded some highest ions. The
model atmospheres allow for departures from LTE for 1127 explicit
levels and superlevels of 46 ions (H, He, C, N, O, Ne, Mg, Al, Si, S,
and Fe), with the same sources for model atoms. Microturbulent velocity is
assumed to be $v_{\rm t} = 2\,{\rm km\,s}^{-1}$.
Recently,  \citet{Dixon+19} have proposed that setting
   microturbulence to zero provides a better fit to the spectra of the
   iron lines in the 21\,400\,K post-AGB star Barnard\,29 in NGC\,6205.
   However, the choice of microturbulence should have little effect on
   the broad hydrogen and helium lines studied here.

\begin{table*}
\caption[]{Parameter space of metal-poor NLTE TLUSTY model spectra}\label{tab:models}
\begin{tabular}{lrrrcrr}
\hline
\hline
Clusters & [M/H] & $T_{\rm eff}$ & $\log g$ & 
$\log {\frac{n_{\rm He}}{n_{\rm H}}}$ & $v_{\rm t}$ & Number of \\
& & [kK] & & &  [km\,s$^{-1}$] & models \\
\hline
NGC\,6121 \& NGC\,6712 & $-$1.0 & 40--70 & 4.00--6.00 & $-$1.5\ldots$+$1.0 & 5. & 256\\
NGC\,5139 \& NGC\,6656 & $-$1.5 & 40--82 & 4.25--6.00 & $-$3.0\ldots$+$0.5 & 5. & 426\\
                      &        & 16--28 & 2.75--5.75 & $-$3.0\ldots\hfill0.0 & 2. & 332\\
NGC\,6779 & $-$2.0 & 18--24 & 2.75--3.25 & $-$1.0\ldots$-$0.5 & 2. & 24\\
NGC\,7099 & $-$2.3 & 50--60 & 5.50--6.75 & $-$3.0\ldots$-$2.0 & 5. & 108\\
\hline
\end{tabular}
\end{table*}

For the cool stars NGC\,5139$-$ROA3596, NGC\,5139$-$UIT1275, and
NGC\,5139$-$UIT1425, which are most likely post-HB stars, we used the
solar metallicity LTE model spectra described in \citet{Moehler+00} to
simulate the effects of radiative levitation. For the re-analysis
  of the stars cooler than 35\,000\,K from \citet{Moehler+98} and for
  NGC\,5139$-$ROA542 we used LTE model spectra like those from
  \citet{Moehler+00} for metallicities [M/H] = $-1.0$ and $-1.5$.
For NGC\,7099$-$UVBS2 we used the metal-free NLTE model spectra
described in \citet{Moehler+98}.

To establish the best fit to the observed spectra, we used the
routines developed by \citet{Bergeron+92} and \citet{Saffer+94}, as
modified by \citet{Napiwotzki+99}, which employ a $\chi^2$ test. The
$\sigma$ necessary for the calculation of $\chi^2$ is estimated from
the noise in the continuum regions of the spectra. The fit program
normalizes model spectra {\em \textup{and}} observed spectra using the same
points for the continuum definition. We fitted the Balmer (and
corresponding He\,{\sc ii}) lines H10 to H$\beta$ (excluding
H$\epsilon$ due to the interstellar Ca\,{\sc ii} absorption line), the
He\,{\sc i} lines $\lambda\lambda$4026.29\,\AA, 4120.81\,\AA,
4387.59\,\AA, 4471.60\,\AA, and 4921.94\,\AA, and the He\,{\sc ii}
lines $\lambda\lambda$4025.60\,\AA, 4199.83\,\AA\ (only for stars
hotter than 30\,000\,K) , 4541.59\,\AA, and 4685.68\,\AA.  These fit
routines underestimate the {\em \textup{formal}} errors by at least a factor of
2 (Napiwotzki priv. comm.). We therefore provide formal errors
multiplied by 2 to account for this effect. In addition, the errors
provided by the fit routine do not include possible systematic errors
due to flat-field inaccuracies or imperfect sky subtraction, for instance.
The results of the line
profile fits are listed in Table\,\ref{tab:params}.

For some of the spectra (NGC\,5139$-$UIT151, NGC\,5139$-$Dk3873,
NGC\,5139$-$ROA5342, and NGC\,6712$-$C49) we could not fit the H,
He\,\,{\sc i}, and He\,\,{\sc ii} lines simultaneously because the
observed He\,\,{\sc i} lines were weaker than predicted, while the H
and He\,{\sc ii} lines were stronger than predicted. We saw similar
effects for NGC\,6121$-$Y453 and NGC\,6723$-$III60 from
\citet{Moehler+98}. All these stars have effective temperatures
between about 50\,000\,K and 63\,000\,K and roughly solar helium
abundances. A similar behaviour was observed by \citet{Latour+15} for
BD$+$28$^\circ$4211. They suspected missing opacities to be the cause of
the problem. They showed that the effects of increasing
metallicity on the hydrogen and helium lines saturate at [M/H] = $+$1
and therefore used model atmospheres with [M/H] = $+$1 for their
analysis. With these model atmospheres, they were able to reconcile the
parameters derived from optical spectra with those derived from UV
spectra. We therefore decided to follow the same path and analysed our
hotter stars with very metal-rich model spectra. For the
synthetic spectra we included only hydrogen and helium lines, as the
high metallicity may only be a proxy for missing opacities (see,
however, Sect.\,\ref{sec:UVphot} and Fig.\,\ref{fig:UV} for further
information).  The new results are listed in
Table\,\ref{tab:metal-rich}. For NGC\,6121$-$Y453 we note that
\cite{Dixon+17} obtained similar parameters from the same optical spectrum,
but a much higher temperature of 72\,000\,K (with $\log g$ of 5.7)
from far-ultraviolet (FUV) data. For reasons of consistency, we use the value
derived here for further discussion.

The effective temperatures clearly decrease and the surface
  gravities increase with the metal-rich model atmospheres. This effect
  is also known for cooler stars (e.g. \citealt{Moehler+00}).
\begin{table*}
\caption[]{Atmospheric parameters for stars hotter than 40\,000\,K
derived using model atmospheres with
  [M/H] = $+$1.}\label{tab:metal-rich}
\begin{tabular}{llrrrr}
\hline
\hline
cluster & star & $T_{\rm eff}$ & $\log g$ & $\log {\frac{n_{\rm He}}{n_{\rm H}}}$ & $M$\\
 & & [K] & [cm s$^{-2}$] & & [M$_\odot$]\\
\hline
NGC\,5139 & UIT151  & 57400$\pm$2200   & 5.37$\pm$0.08& $-$1.07$\pm$0.08 & 0.51\\
          & Dk3873  & 47900$\pm$1200& 5.15$\pm$0.08& $-$1.19$\pm$0.08 & 0.42\\
          & ROA5342 & 49800$\pm$1400& 4.98$\pm$0.08& $-$1.15$\pm$0.06 & 0.46\\[1mm]
NGC\,6121 & UVBS2   & 46200$\pm$\ \ 400& 5.92$\pm$0.12& $+$0.96$\pm$0.14 & 1.06\\
NGC\,6656 & UVBS2   & 62300$\pm$9600   & 4.59$\pm$0.16& $+$0.40$\pm$0.18 & 0.31\\
NGC\,6712 & C49     & 47800$\pm$1800& 4.64$\pm$0.08& $-$1.10$\pm$0.08 & 0.51\\
\hline
\multicolumn{5}{c}{New analysis of data from \cite{Moehler+98}}\\
\hline
NGC\,6121 & Y453    & 56500$\pm$1800 & 5.71$\pm$0.12 & $-$1.16$\pm$0.10 & 0.79 \\
NGC\,6723 & III60   & 42300$\pm$1400 & 4.80$\pm$0.12 & $-$1.13$\pm$0.12 & 1.19\\
\hline
\end{tabular}
\end{table*}

\subsection{NGC\,6656$-$UVBS2}
The analysis of NGC\,6656$-$UVBS2 was difficult for two reasons: It
has a cool neighbour very close by, and it has a planetary nebula. The
first results using model spectra with the cluster abundance pointed
towards a temperature of almost 80\,000\,K, while the results obtained
with metal-rich model spectra suggest a much cooler temperature of
62\,000\,K. Based on a low-resolution spectrum, \cite{HaPa93} suggested
an effective temperature of 75\,000\,K from the similarity to the sdO
star KS\,292, which had been analysed by \cite{Rauch+91}. The spectrum
shows absorption lines of C\,{\sc iv} and N\,{\sc v} similar to the
star SDSS\,J155610.40$+$254640.3 \citep{Reindl+16}, which has a
temperature of about 100\,000\,K.  The deeper He\,{\sc ii} lines of
NGC\,6656$-$UVBS2 are consistent with a lower temperature, which
unfortunately does not allow us to distinguish between the two
temperatures mentioned before. 

\citet{Muthu+13}
suggested that NGC\,6656$-$UVBS2 is the result of a stellar merger.
Interestingly enough, SDSS\,J155610.40$+$254640.3
  was classified by \citet{Reindl+16} as a PG\,1159 star, a class of
  H-deficient post-AGB stars. In any case, it seems that
  NGC\,6656$-$UVBS2 is closely related to the star ZNG1 in NGC\,5904
  (see Table\,\ref{tab:lit}), which has also been connected to both a
  merger and PG\,1159 stars \citep{Dixon+04}. While
    NGC\,5904$-$ZNG1 shows a high rotational velocity of about
    170\,km s$^{-1}$ , we see no evidence for fast rotation in our
    spectrum of NGC\,6656$-$UVBS2. We recall, however,
    that the resolution of our data is about 100\,km s$^{-1}$.
\subsection{Metal lines}\label{sec:metal}
We took care to verify that
potential metal lines were present in both individual spectra for each
star. We found potential metal lines only in the stars listed
below and compared their strength to that predicted by the best-fitting model
spectrum. The number of lines for each ion are given in parentheses,
and the stars are sorted by evolutionary stage. A detailed abundance
analysis is beyond the scope of this paper.

\begin{description}
\item [NGC\,5139$-$UIT1425 (pHB)] Mg\,{\sc ii} (1) significantly
    weaker than predicted by the metal-rich model spectrum with metal
    lines, which is usual for HB stars in this temperature range.
\item [NGC\,5139$-$UIT151 (peAGB)] N\,{\sc v} (2) in agreement with the
  metal-rich model spectrum with metal lines, which also
    predicts strong C\,{\sc iii} and O\,{\sc iv} lines that are not
    observed, however.
\item [NGC\,5139$-$UIT644 (pAGB)] N\,{\sc ii} (9), C\,{\sc ii} (1)
Mg\,{\sc ii} (1), and Si\,{\sc iii} (2) much stronger than predicted
by the metal-poor model spectrum used for the analysis, especially
N\,{\sc ii}. 
\item [NGC\,6656$-$UVBS2 (pAGB)] C\,{\sc iv} (3), N\,{\sc v} (2),
  much stronger than predicted by the model spectrum with
  metal lines for [M/H] = $+$1.
\item [NGC\,6121$-$UVBS2 (WDM)] C\,{\sc iii} (7), 
  much stronger than predicted by the model spectrum with
  metal lines for [M/H] = $+$1.
\end{description}
\subsection{UV spectrophotometry}\label{sec:UVphot}
\begin{figure*}
\includegraphics[height=\textwidth, angle=270]{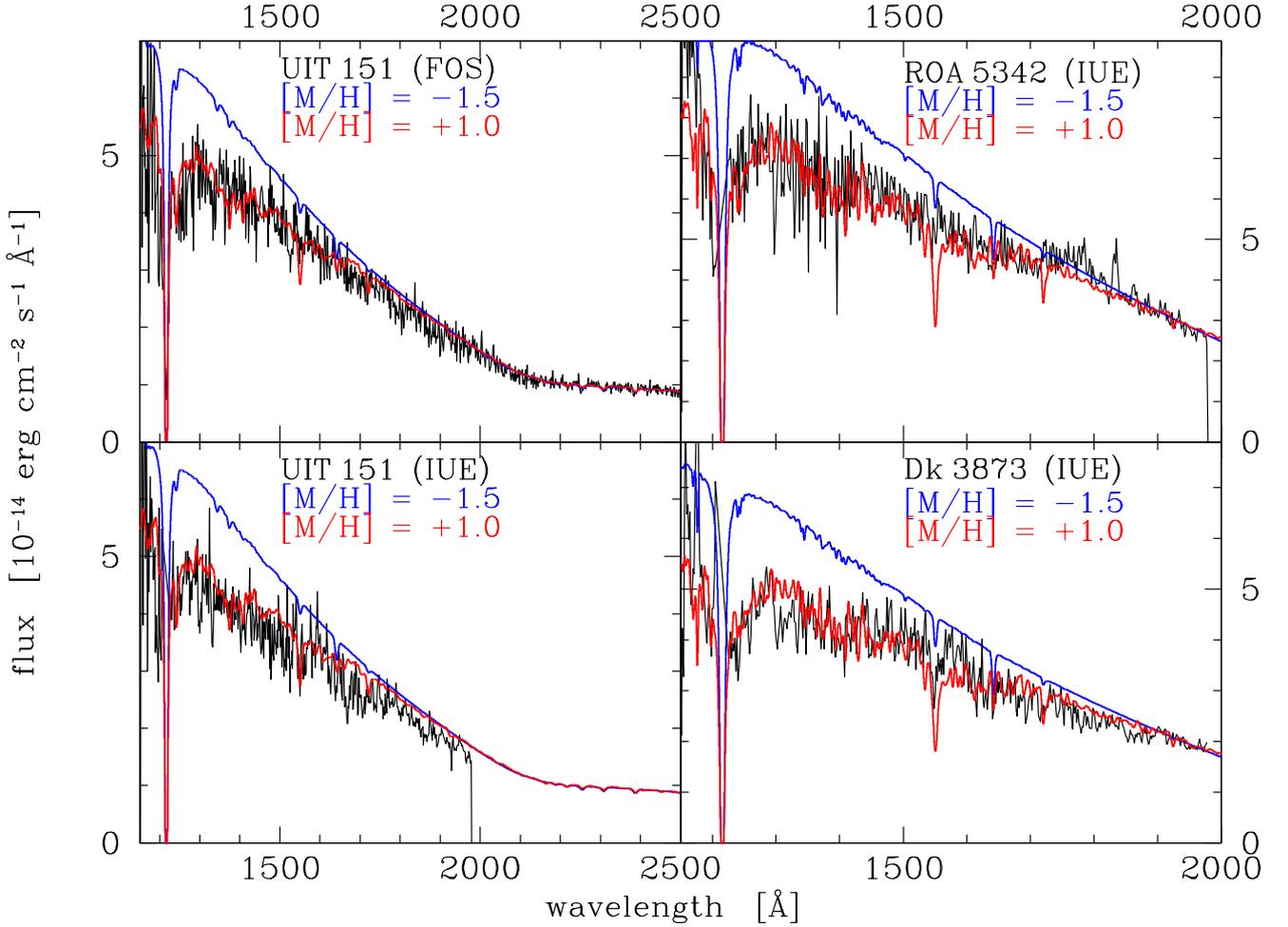}
\caption[]{FOS (UIT151, top left) and IUE (UIT151, ROA5342, and
  Dk3873) spectra together with model spectra for [M/H] = $-$1.5 (blue) and
  [M/H] = $+$1.0 (red). The model spectra were calculated for the
  temperature, surface gravity, and helium abundance listed in
  Table\,\ref{tab:metal-rich}. They were reddened by $E_{B-V}$ = 0.13
  and aligned to a radial velocity of $+$232 km\,s$^{-1}$. The model
  spectra were scaled to the observed flux between 2400\,\AA\ and 2500\,\AA\
  in the FOS spectrum and between 1900\,\AA\ and
  1950\,\AA\ in the IUE spectra. Metal-rich model spectra are clearly needed to reproduce
  the observed UV flux.}\label{fig:UV}
\end{figure*}

Several of our targets have archival UV spectra from either
the {International Ultraviolet Explorer ({\bf IUE})}, the {Far Ultraviolet
  Spectroscopic Explorer ({\bf FUSE}),} or the HST.  We defer a complete
discussion of the UV spectra to a subsequent paper, but
discuss here some of the IUE and HST Faint Object Spectrograph ({\bf FOS})
data in $\omega$\,Cen (Table\,\ref{tab:uvspectra}), which (1) provide
additional justification for our use of metal-rich atmospheres for the
sdO stars, and (2) provide evidence of binarity for the star UIT1425.
The data were obtained from the
\href{http://archive.stsci.edu/iue/}{Mikulski Archive for Space
  Telescope ({\bf MAST})}, and the IUE spectra were converted into the HST
absolute calibration scale using the transformation in \citet{BoBi18}.
We restricted use of the IUE data to the short-wavelength ($<$ 1950 \AA)
prime ({\bf SWP}) camera because the IUE large (10\arcsec$\times$20\arcsec)
aperture includes light from background red stars at longer
wavelengths.  We expect the SWP images to be free of background
contamination because examination of the UIT 1620\,\AA\ image of
$\omega$\,Cen showed that no UV sources are included within the IUE aperture
for any of the target stars.  In addition, the FOS and IUE spectra of
UIT151 show fair agreement in Fig.\,\ref{fig:UV}, even though the FOS data
use a much smaller (1\arcsec\ circular) aperture.

\begin{table*}
  \caption{Ultraviolet spectra of stars in $\omega$\,Cen}\label{tab:uvspectra}
\begin{tabular}{llrllrrlrr}
\hline
\hline
Star &  Telescope & Mode  & ID & Date & Exp \\
     &            &       &    &      & [s]\\
\hline
UIT151 & HST/FOS & G160L & Y2SS402T & 1996-04-28 &   2400 \\
UIT151 & IUE           & SWP & 54154 & 1995-03-16 & 23400 \\
Dk3873 & IUE  & SWP & 48271 & 1993-07-31 & 19200 \\
ROA5342 & IUE & SWP & 54333 &  1995-04-08    & 25500 \\
UIT1425 & IUE & SWP & 54804 & 1995-05-31 & 26820 \\
\hline
\end{tabular}
\end{table*}

The high-resolution study of field sdO stars by \citet{Latour+18}
showed that the UV spectrum of these stars is dominated by numerous
lines of iron and nickel.  Even though our low-resolution UV spectra
cannot individually resolve these lines, the
lines are sufficiently numerous at high metallicity to affect the apparent continuum.
Figure\,\ref{fig:UV} shows surprisingly good agreement between the
low-resolution UV spectra of the three sdO stars and models with [M/H]
= $+$1.0. The C\,{\sc iv} doublet near 1550\,\AA\ appears much weaker
than predicted by the metal-rich models, suggesting that the light
elements do not follow the abundance enhancement seen in iron and
nickel, an effect also seen in some field sdO stars \citep{Latour+18}.
We are able to use the known metallicity spread of $\omega$ Cen to set
an upper limit of [M/H] = $-$1.0 for the metallicity of the
progenitors of these sdO stars.  The factor of $\approx$100 enhancement
in iron and nickel must then be attributed to radiative levitation
\citep{Latour+18}.

\begin{figure}
\includegraphics[height=\columnwidth, angle=270]{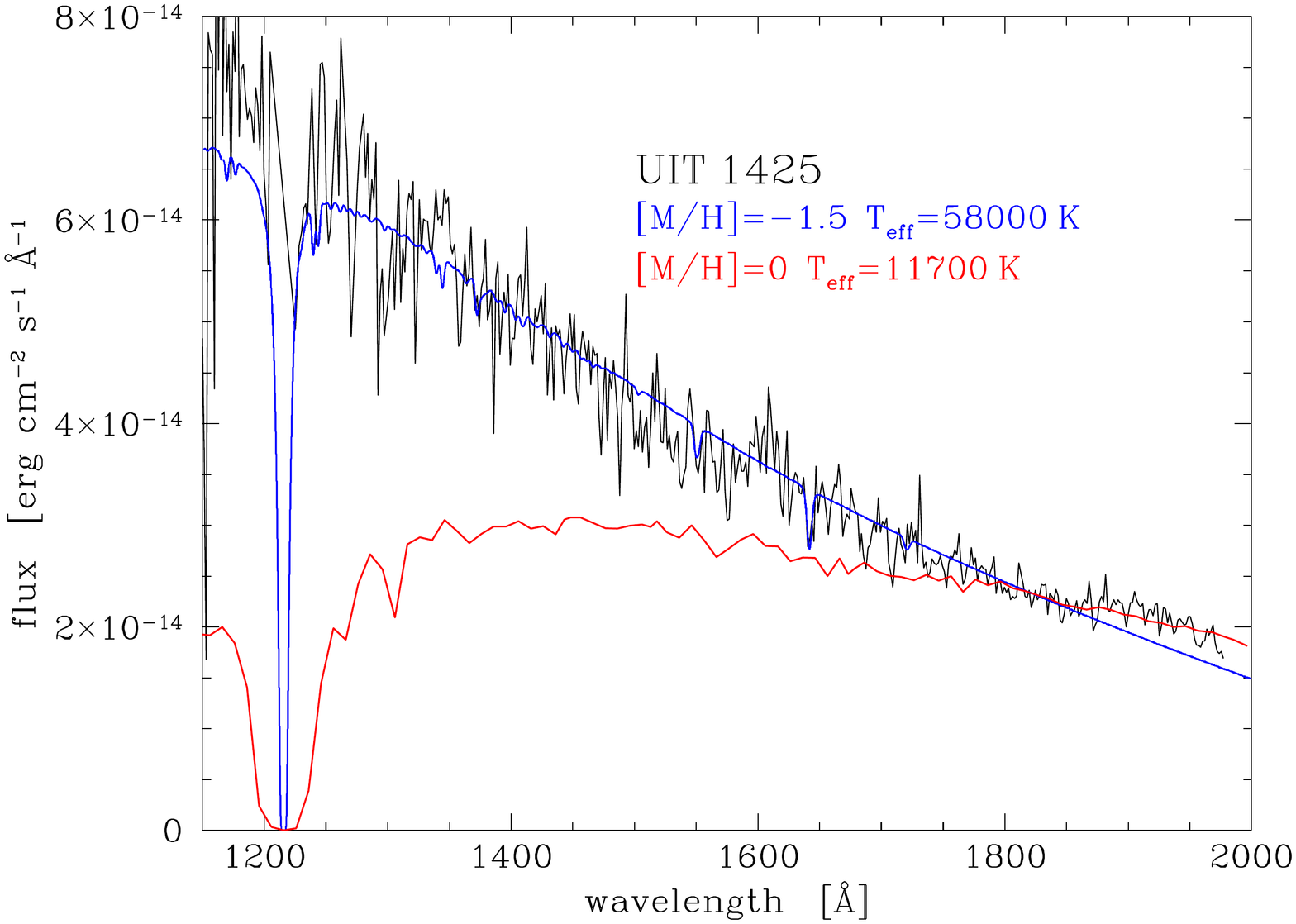}
\caption[]{IUE spectrum for NGC\,5139$-$UIT1425 together with model
  spectra for [M/H] = $-$1.5 (blue, $T_{\rm eff}$ = 58\,000\,K) and [M/H]
  = 0 (red, $T_{\rm eff}$ = 11\,700\,K). The model spectra were reddened
  by $E_{B-V}$ = 0.13, aligned to a radial velocity of $+$232 km
  s$^{-1}$, and scaled to the observed flux between 1800\,\AA\, and
  1900\,\AA. A hot star is clearly needed to reproduce
  the observed UV flux.}\label{fig:UIT1425}
\end{figure}

For NGC\,5139$-$UIT1425 a model spectrum with the parameters
  derived from our optical spectra cannot reproduce the UV data (see
Fig.\,\ref{fig:UIT1425}). This star is most likely a binary, as its
optical colours ($B-V$=$-$0.01, $V-I$=$+$0.20) are appropriate for a
blue HB star, whereas its $UV-V$ colour and steep IUE spectrum is more
indicative of an sdO star. The optical spectrum, however,
shows no trace of He\,{\sc ii}, which would be a clear sign of the
hotter star.  Unfortunately, there is no HST imagery in this region of
$\omega$\,Cen, so that we cannot rule out that the binary is a chance
alignment.  We checked the spatial (line by line) IUE image, but found
no presence of more than one star within the large aperture.

\section{Data from the literature}\label{sec:lit}
In Table\,\ref{tab:lit} we list the physical parameters published for
hot UV-bright stars other than those discussed here. For
NGC\,6656$-$II-81 we note that its position in the $T_{\rm eff}, \log g$
diagram is consistent with being a post-AGB instead of a post-HB
star. 
\begin{table*}
  \caption{Atmospheric parameters of hot UV-bright stars from refereed
    publications until 2019-03-01. The masses were derived by us as
    described in Sect.\,\ref{sec:masses}.}\label{tab:lit}
\begin{tabular}{llrllrrlrr}
\hline
\hline
cluster & star & $T_{\rm eff}$ & $\log g$ & $\log {\frac{n_{\rm He}}{n_{\rm H}}}$
& status & Ref. & $V$ & Ref. & $M$\\
 & & [K] & [cgs] & & & & [mag] & & [M$_\odot$]\\
\hline
NGC\,104   & BS &   11000 & 2.5 & $-$1 & pAGB & (1) & 10.73 & (13)& 1.25\\
NGC\,1851 & UV5  & 16000 & 2.5 & $-$1 & pAGB & (2) & 13.26& (14)& 0.41\\
NGC\,5139 & ROA5701 & 25000 & 3.3 & $-$1 & pAGB & (3) & 13.13 & (24) & 0.32\\
NGC\,5272 & vZ1128 & 36600 & 3.95 & $-$0.84 & peAGB & (4) & 15.03 & (15)& 0.41\\
NGC\,5904 & ZNG1 & 44300 & 4.3 & +0.52 & pAGB & (5) & 14.54 & (16)& 0.72\\
NGC\,5986 & ID6 & 8750 & 2.0 & $-$1.4 & pAGB & (12) & 12.65 & (22) & 1.21\\
NGC\,6205 & B29   & 21400 & 3.10 & $-$0.89 & pAGB & (6) & 13.116& (23) & 0.38\\
NGC\,6254 & ZNG1 & 27000 & 3.6 & $-$1.33 & pAGB & (7) & 13.23 & (17)& 0.59\\
NGC\,6397 & ROB162 & 51000 & 4.5 & $-$1.0 & pAGB & (8) & 13.1 & (9)& 0.50\\
NGC\,6656 & II-81 & 38000 & 4.2 & --- & pHB & (9) & 14.0 & (18) & 0.47\\
NGC\,6712 & ZNG1 & 11000 & 2.1 &  $-$1.19 & pAGB & (7) & 13.33 & (19)& 0.35\\ 
NGC\,6752 & B852  & 39000 & 5.2 & $-$2.0 & pHB & (10) & 15.91 & (20)& 0.52\\
          & B4380 & 32000 & 5.3 & $-$2.3   & pHB & (10) & 15.93 &(20)& 0.79\\
          & B1754 & 40000 & 5.0 & $-$1.52  & pHB & (10) & 15.99 & (20)& 0.30\\
NGC\,7078 & K648  & 39000 & 3.9  & $-$1.10 & pAGB & (11) & 14.73 & (11)& 0.62\\
          & K996  & 11500 & 2.5 & --- & pHB & (12) & 14.31 & (21)& 0.26\\ 
          & ZNG1  & 28000 & 3.7  & $-$1.22 & pAGB & (7) & 14.8 & (17)& 0.55\\
\hline
\end{tabular}
\tablebib{
(1) \cite{Dixon+95}; 
(2) \cite{Dixon+94}; 
(3) \cite{Thompson+07}; 
(4) \cite{Chayer+15}; 
(5)  \cite{Dixon+04}; 
(6) \cite{Dixon+19}; 
(7) \cite{Mooney+04}; 
(8) \cite{HeKu86}; 
(9) \cite{Glaspey+85}; 
(10) \cite{Moehler+97}; 
(11) \cite{Rauch+02}; 
(12) \cite{Jasniewicz+04}. 
(13) \cite{LloydEvans74} 
(14) \cite{Walker92} 
(15) \cite{Buzzoni+92} 
(16) \cite{Piotto+02} 
(17) \cite{deBoer87} 
(18) \cite{ArpMelbourne59} 
(19) \cite{Cudworth88} 
(20) \cite{Buonanno+86} 
(21) \cite{Buonanno+83} 
(22) \cite{Alves+01} 
(23) \cite{Sandquist+10} 
(24) \citet{Bellini09}
}
\end{table*}

We wish to compare results between different globular clusters, therefore we
decided to use homogeneous data sets for their ages
\citep{MarinFranch+09}, metallicities \citep{Carretta+09}, and
integrated $V$ magnitudes \citep{vandenBergh91}. Unfortunately, we
found no such data set for the distances and reddenings. Therefore we
used the values from \citet{Harris96} that are at least created in a
defined way from the various sources. The values we used in our
  analysis are listed in Table\,\ref{tab:par-gc}.
\begin{table*}
  \caption{Metallicities, ages, integrated $V$ magnitudes, distance
    moduli, reddenings, and bolometric corrections for the globular clusters listed in
    Tables\,\ref{tab:params} and \ref{tab:lit}.}\label{tab:par-gc} 
\begin{tabular}{l|rlrrrr|rrr}
  \hline
  \hline
Cluster & [M/H]\tablefootmark{a} & age\tablefootmark{b} & $V_t$\tablefootmark{c} & $(m-M)_V$\tablefootmark{d} & $E_{B-V}$\tablefootmark{d} & $BC_V$\tablefootmark{e} & \multicolumn{3}{c}{Fraction of HB stars\tablefootmark{f} }   \\
  & &    [10$^9$ yr]  & [mag]  & [mag]    & [mag]       & [mag]    & $f_{\rm RHB}$ & $f_{\rm BHB}$ & $f_{\rm EHB}$ \\\hline
  NGC\,104  & $-$0.76 & 13.1 & 4.01 & 13.37 & 0.04 & $-$0.53 & 1.00 & 0.00 & 0.00 \\ 
  NGC\,1851 & $-$1.18 & 10.0 & 7.16 & 15.47 & 0.02 & $-$0.41 & 0.64 & 0.36 & 0.00 \\ 
  NGC\,5139 & $-$1.64 & 11.5 & 3.85 & 13.94 & 0.12 & $-$0.38 & 0.06 & 0.71 & 0.23 \\ 
  NGC\,5272 & $-$1.50 & 11.4 & 6.36 & 15.07 & 0.01 & $-$0.38 & 0.32 & 0.68 & 0.00 \\ 
  NGC\,5904 & $-$1.33 & 10.6 & 5.58 & 14.46 & 0.03 & $-$0.40 & 0.21 & 0.79 & 0.00 \\ 
  NGC\,5986 & $-$1.63 & 12.2 & 7.46 & 15.96 & 0.28 & $-$0.38 & 0.00 & 0.91 & 0.09     \\   
  NGC\,6121 & $-$1.18 & 12.5 & 5.77 & 12.82 & 0.35 & $-$0.43 & 0.39 & 0.61 & 0.00 \\ 
  NGC\,6205 & $-$1.58 & 11.6 & 5.82 & 14.33 & 0.02 & $-$0.38 & 0.00 & 0.70 & 0.30 \\ 
  NGC\,6254 & $-$1.57 & 11.4 & 6.60 & 14.08 & 0.28 & $-$0.38 & 0.00 & 1.00 & 0.00 \\ 
  NGC\,6397 & $-$1.99 & 12.7 & 6.20 & 12.37 & 0.18 & $-$0.37 & 0.00 & 1.00 & 0.00  \\ 
  NGC\,6656 & $-$1.70 & 12.7 & 5.08 & 13.60 & 0.34 & $-$0.38 & 0.00 & 0.73 & 0.27 \\ 
  NGC\,6712 & $-$1.02 & 12.0 & 8.04 & 15.60 & 0.45 & $-$0.44 & 0.85 & 0.15 & 0.00 \\ 
  NGC\,6723 & $-$1.10 & 13.1 & 7.03 & 14.84 & 0.05 & $-$0.44 & 0.58 & 0.42 & 0.00 \\ 
  NGC\,6752 & $-$1.55 & 11.8 & 5.32 & 13.13 & 0.04 & $-$0.38 & 0.00 & 0.72 & 0.28 \\ 
  NGC\,6779 & $-$2.00 & 13.7 & 8.17 & 15.68 & 0.26 & $-$0.37 & 0.00 & 1.00 & 0.00 \\ 
  NGC\,7078 & $-$2.33 & 12.9 & 6.32 & 15.39 & 0.10 & $-$0.35 & 0.10 & 0.90 & 0.00 \\ 
  NGC\,7099 & $-$2.33 & 12.9 & 7.35 & 14.64 & 0.03 & $-$0.35 & 0.00 & 1.00 & 0.00 \\ 
\hline
\end{tabular}
\tablefoot{
\tablefoottext{a}{\cite{Carretta+09}}
\tablefoottext{b}{\cite[for NGC\,6712 the age is taken from
    \citet{Paltrinieri+01}]{MarinFranch+09}}
\tablefoottext{c}{\cite{vandenBergh91}}
\tablefoottext{d}{\cite[2010 edition]{Harris96}}
\tablefoottext{e}{\cite{Worthey1994}}
  \tablefoottext{f}{The fraction of RHB, BHB, and EHB stars populating
    the HB ($f_{\rm RHB}$, $f_{\rm BHB}$, and $f_{\rm EHB}$,
    respectively) of each cluster have been derived from the The
    Hubble Space Telescope UV Legacy Survey of Galactic Globular
    Clusters (\citealt{Piotto+15,Soto+17}), with the exception of
    $\omega$\,Cen and NGC\,6712, which are not part of the survey. For
    NGC\,6712, which lacks an EHB, the RHB/BHB fractions were derived
    from the HB ratio presented by \cite{Paltrinieri+01}, while the
    EHB, BHB, and RHB fractions for  $\omega$\,Cen were derived from
    \citet{Castellani+07} and \citet{Calamida+17}. }
  }
\end{table*}

\section{Masses and distances}\label{sec:masses}
\begin{figure}
\begin{center}
\includegraphics[height=\columnwidth, angle=270]{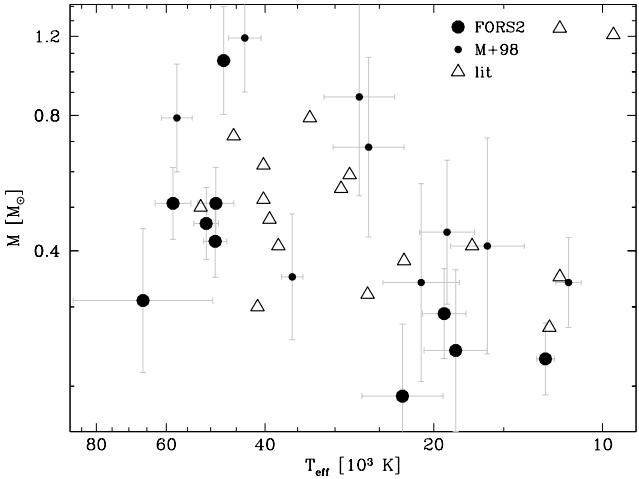}
\caption[]{Effective temperatures and masses for the hot UV-bright stars in globular clusters. For stars hotter than 40\,000\,K,
  we show the results obtained with model spectra with [M/H] =
  $+$1. } \label{fig:all_tm}
\end{center}
\end{figure}

Using the atmospheric parameters together with the observed
  brightness of the stars and reddening and distances to their parent
  globular cluster, we determined masses using Eq.\,\ref{eq:mass_bc},
\begin{equation}\label{eq:mass_bc}
 \log {\frac{M}{M_\odot}} = \log{\frac{g_\ast}{g_\odot}} - 4\cdot \log
 {\frac{T_\ast}{T_\odot}} - {\frac{M_V+BC-4.74}{2.5}}
,\end{equation}
which can be rewritten to
\begin{equation}\label{eq:mass_bc2}
 \log {\frac{M}{M_\odot}} =  \log g_\ast + 0.4 \cdot
 [(m-M)_0-V_\ast+A_V - BC] -4 \cdot \log T_\ast +C3
,\end{equation}
with
\begin{equation}\label{eq:C3}
 C3 = -\log g_\odot + 4\cdot \log T_\odot + {\frac{4.74}{2.5}}
,\end{equation}

with the bolometric corrections from \citet{Flower1996}. The results
are listed in Tables\,\ref{tab:params} and \ref{tab:lit} and plotted in
Fig.\,\ref{fig:all_tm}.

From stellar evolution we would generally expect masses of
0.5\,M$_\odot$ to 0.6\,M$_\odot$. Figure \,\ref{fig:all_tm} shows that
the masses derived for stars below 25\,000\,K
tend to be too low, while the masses derived for hotter stars scatter
towards too high masses\footnote{The dust in the planetary
  nebula IRAS\,18333$-$2357 surrounding NGC\,6656$-$UVBS2 (at about
  62\,000\,K) probably provides additional extinction so
  that the derived mass is definitely a lower limit.  \citet{HaPa93}
  estimated a reddening of $E_{B-V}$ = 0.54 compared to a global
  reddening of 0.34 used for the mass determination. This would
  increase the mass by some 80\%, leading to a mass of
  $\sim 0.55\,M_\odot$, which is typical of low-mass central stars of
  planetary nebulae.}. 

  Some of the globular clusters discussed in this paper are close
  enough to have parallax estimates from the {\it Gaia} DR2
  \citep{Gaia}. The Gaia collaboration compared the parallaxes derived from the
  {\it Gaia} measurements to those listed in \citet{Harris96} and
  found a systematic difference of 0.029\,mas, with the {\it Gaia}
  parallaxes being smaller. We therefore used the reported {\it Gaia}
  parallaxes only for those clusters that had parallax values 
  of more than 0.23\,mas, that is, about eight times the systematic offset. To
  derive distances, we applied the reported systematic offset to the
  parallaxes. We then added interstellar extinction as
  3.2$\cdot {E_{B-V}}$. The apparent distance moduli derived
  that way are 0.29\,mag smaller for NGC\,6121 and 0.16\,mag,
  0.18\,mag, and 0.08\,mag larger for NGC\,6397, NGC\,6656, and
  NGC\,6752, respectively, than those from \citet{Harris96}. It is
  interesting to note that we find higher-than-expected masses for
  the two stars in NGC\,6121, which would be reduced by 30\% with the
  {\it Gaia} distance to 0.81\,M$_\odot$ and 0.65\,M$_\odot$,
    respectively. For NGC\,6397,  using the {\it Gaia} distance would
  increase the masses by 16\%, moving them from about 0.5\,M$_\odot$
  to about 0.6\,M$_\odot$. For NGC\,6656, using the {\it Gaia} distance
  would increase the mass of NGC\,6656$-$UVBS2 from 0.31\,M$_\odot$ to
  0.38\,M$_\odot$ (without the additional reddening correction) and
  from 0.55\,M$_\odot$ to 0.65\,M$_\odot$ (with the additional
  reddening correction). 
The change of 8\% for the masses of the stars in
  NGC\,6752 is negligible.

\section{Stellar evolution models and evolutionary fluxes}
\label{sec:evol}
\begin{table}
\caption{Initial parameters of the stellar evolution models}\label{tab:par-gc-model}
\label{tab:evol_ini}
\begin{tabular}{lllll}
\hline
\hline
[M/H] & $M_{\rm ZAMS}$  & $Z_{\rm ZAMS}$ & $Y_{\rm ZAMS}$ & $X_{\rm ZAMS}$\\
 & [M$_\odot$] & & & \\
\hline
$-$1.0  &  0.85 & 0.00172   & 0.24844  & 0.74984 \\
$-$1.5  &  0.83 & 0.000548  & 0.246096 & 0.753356\\
$-$2.0  &  0.82 & 0.000174  & 0.245348 & 0.754478\\
$-$2.3  &  0.82 & 0.000087  & 0.245174 & 0.754739\\
\hline
\end{tabular}
\end{table}

\begin{table}
  \caption{Number of expected hot post-AGB UV-bright stars with $\log
    L/L_\odot >2.65$ and $80\,000\,K > T_{\rm eff} > 7000$\,K (N$^{\rm
      post-AGB}_{\rm exp}$) for the various clusters
     compared with the actual number of post-AGB stars in each
    cluster (N$^{\rm post-AGB}_{\rm obs}$).  }
\label{tab:result-gc}
\begin{tabular}{lc|ccc}
\hline
\hline
 Cluster & N$^{\rm post-AGB}_{\rm obs}$  &   [M/H]$_{\rm tracks}$ &  
N$^{\rm post-AGB}_{\rm exp}$\\
\hline
 NGC\,104  &    1  &   $-$1.00   &     0.08   -- \hfill   0.65     \\
NGC\,1851  &    1  &   $-$1.00   &     0.15   -- \hfill   1.04    \\
NGC\,5139  &    5  &   $-$1.50   &     1.31   -- \hfill  13.58     \\
NGC\,5272  &    1  &   $-$1.50   &     0.27   -- \hfill   1.48     \\
NGC\,5904  &    1  &   $-$1.50   &     0.38   -- \hfill   2.11     \\
NGC\,5986  &    1  &   $-$1.50   &     0.61 -- \hfill 4.29             \\
NGC\,6121  &    1  &   $-$1.00   &     0.19   -- \hfill   1.32     \\
NGC\,6205  &    1  &   $-$1.50   &     0.23   -- \hfill   2.71     \\
NGC\,6254  &    1  &   $-$1.50   &     0.26   -- \hfill   1.44     \\
NGC\,6397  &    1  &   $-$2.00   &     0.07   -- \hfill   0.26    \\
NGC\,6656  &    2  &   $-$1.50   &     0.59   -- \hfill   6.65     \\
NGC\,6712  &    2  &   $-$1.00   &     0.14 -- \hfill 0.98    \\
NGC\,6723  &    2  &   $-$1.00   &     0.12   -- \hfill   0.83     \\
NGC\,6752  &    0  &   $-$1.50   &     0.13   -- \hfill   1.48     \\
NGC\,6779  &    1  &   $-$2.00   &     0.29   -- \hfill   1.13     \\
NGC\,7078  &    2  &   $-$2.30   &     1.04   -- \hfill   3.71     \\
NGC\,7099  &    0  &   $-$2.30   &     0.17   -- \hfill   0.64     \\
\hline
\end{tabular}
\end{table}

\begin{figure*}
\begin{center}
\includegraphics[height=\textwidth, angle=270]{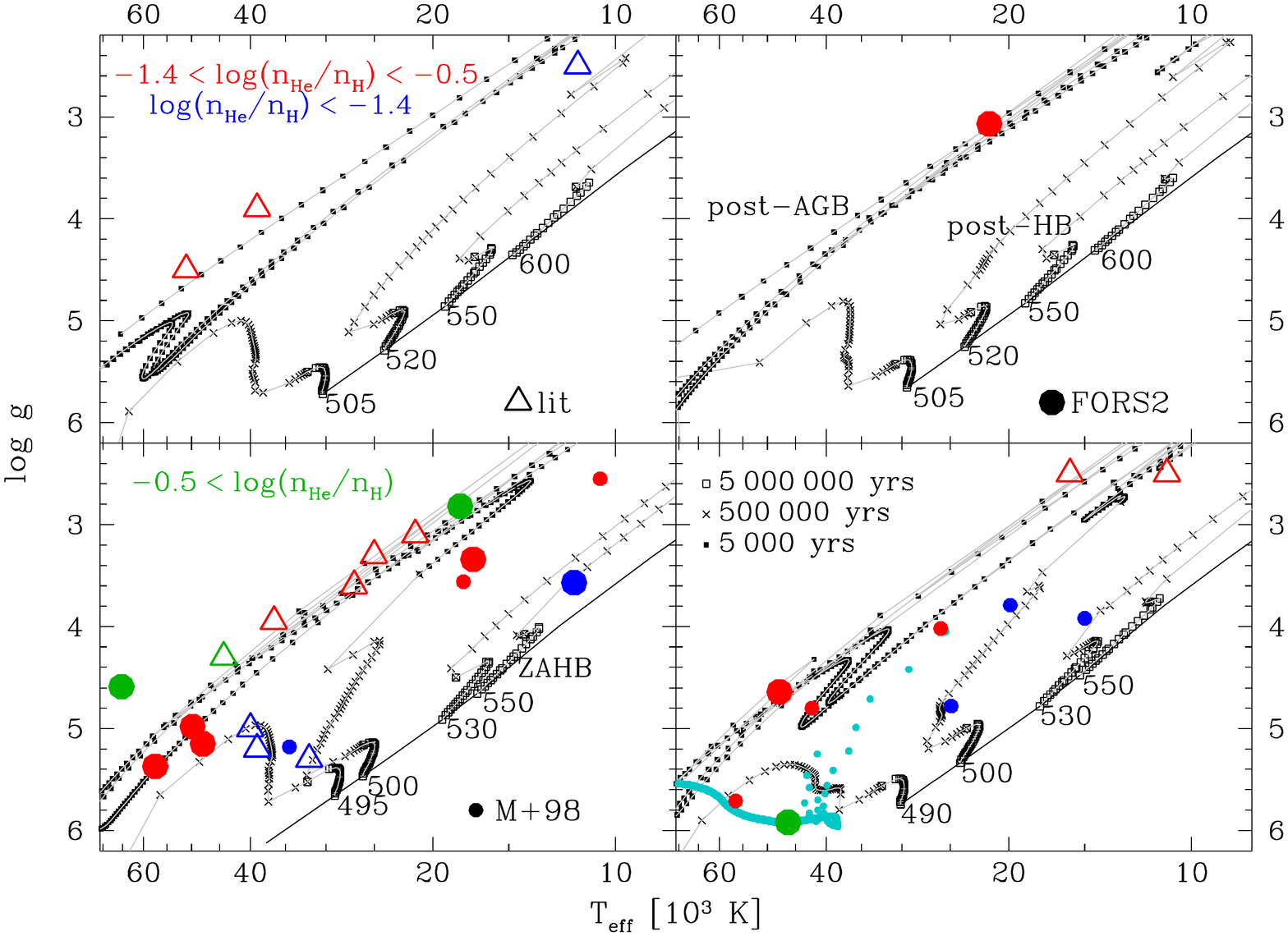}
\caption[]{Effective temperatures and surface gravities for the hot UV
-bright stars in globular clusters. For stars hotter than 40\,000\,K,
  we show the results obtained with model spectra with [M/H] =
  $+$1. Stars from the literature without helium abundance are assumed
to have solar abundance.
The panels show the metallicity bins $-$2.3 (top left), $-$2
  (top right), $-$1.5 (bottom left), and $-$1.0 (bottom right). The
  symbols along the tracks mark constant time steps. The cyan symbols mark
  time steps of 50\,000 years along the white dwarf merger track from
  \citet{Zhang+12}. }\label{fig:all_tg}
\end{center}
\end{figure*}
The hot UV-bright phase contains stars in different evolutionary stages. 
The most luminous of these stars ((L/L$_\odot$)\ $\gtrsim$ 3.1) 
are believed to be post-AGB stars,
which go through a luminous UV-bright phase as they leave the AGB and
move rapidly toward their final white dwarf state.  Despite their
short lifetimes ($\lesssim 10^{5}$\,yr), hot post-AGB stars
($T_{\rm eff} >$ 7\,000\,K) can dominate the total UV flux
of an old stellar population.  
UV-bright stars with luminosities 2.65~$\lesssim \log$~(L/L$_\odot$)\
$\lesssim$~3.1 are a mixture of low-mass post-AGB stars, which are stars that
departed from the AGB before the beginning of the thermal pulses
(post-early AGB stars), and post-EHB stars that managed to ignite the
H-burning shell after the departure from the HB, but never reached the
AGB. The post-early AGB population arises from hot HB stars with
sufficient envelope mass to return to the AGB, but which peel off the
AGB prior to the thermally pulsing phase \citep{Dorman+93}.
Less luminous hot UV-bright stars (1.8~$\lesssim \log$~(L/L$_\odot$)\
$\lesssim$ 2.65) are mostly evolving from the EHB to the white dwarf
domain (post-EHB stars), but some may also evolve from the hot end of
the blue horizontal branch ({\bf BHB}) to the AGB phase. The
population of post-EHB stars is expected to be about 15--20\% of the
population of EHB stars \citep{Dorman+93}.

We now wish to compare the number of observed post-AGB stars with
expectations from stellar evolution models.  In the absence of a
detailed study of the distribution of stars in the HB
of each cluster it is not possible to perform detailed synthetic
population simulations for each cluster. However, using the available
information from the HST UV Legacy Survey of Galactic Globular
Clusters \citep{Piotto+15,Soto+17} we estimated
the fractions of stars populating the red HB ({\bf RHB}), BHB, and
EHB, based on the HB morphology in
  the $V, V-I$ colour-magnitude diagram (see Table\,\ref{tab:par-gc}
    for results). Using these numbers, we estimated
how many post-AGB stars should evolve from the RHB, BHB, and EHB in
each cluster. 

The post-AGB evolutionary models computed for this work are an
extension of the models presented by \cite{MillerBertolami2016}. The
models were computed for values of [M/H]$=-1$, $-1.5$, $-2$ and $-2.3$
under the assumption of a scaled-solar mixture
\citep{Grevesse+98}. The helium-metallicity relation was taken as in
\cite{MillerBertolami2016}, that is, $Y=0.245+2\times Z$. The initial
masses of the models were chosen to correspond to ages between 11.5
to 12 Gyr. The higher or lower ages of some of the globular clusters
(see Table\,\ref{tab:par-gc}) are no problem because age does not influence
the mass of the helium core at the time of the helium core flash, nor
the following evolution. It might affect the mass of the hydrogen-rich
envelope, but this mass is varied artifically in our models anyway to
populate the whole horizontal branch. The corresponding initial
parameters for the stellar evolution sequences are shown in
Table\,\ref{tab:evol_ini}. Different mass loss was applied to each
model before the zero-age horizontal branch ({\bf ZAHB}) in order to
obtain different masses at the ZAHB and a complete coverage of the
horizontal branch (from the RHB to the EHB) and post-HB evolution (see
Table\,\ref{tab:sequences} for details). Convective boundary mixing at
all evolutionary stages and mass loss on the AGB were also adopted as
in \cite{MillerBertolami2016}. Figure\,\ref{fig:all_tg} shows the
atmospheric parameters derived in Sect.\,\ref{sec:param} compared to
the evolutionary tracks. The evolutionary stage derived from this
comparison can be found in Table\,\ref{tab:params}. The stars are
  colour-coded by their helium abundance because this helps to identify
  the evolutionary status of a star. Significantly sub-solar helium
  abundances due to diffusion are known for HB stars
  hotter than about 11\,000\,K. The diffusion patterns are erased once
  convection starts in the atmosphere as the star evolves towards
  cooler temperatures. A low helium abundance is therefore unlikely
  for stars that ascended the AGB (even if only
  partially). Fig.\,\ref{fig:all_tg} shows that helium-poor stars are
  found along post-(E)HB tracks and not along post-(e)AGB
  tracks.  Helium-rich stars, on the other hand, are found only
  along the  post-AGB tracks (except for the white dwarf merger). 

\begin{figure*}
\begin{center}
\includegraphics[width=\textwidth, angle=0]{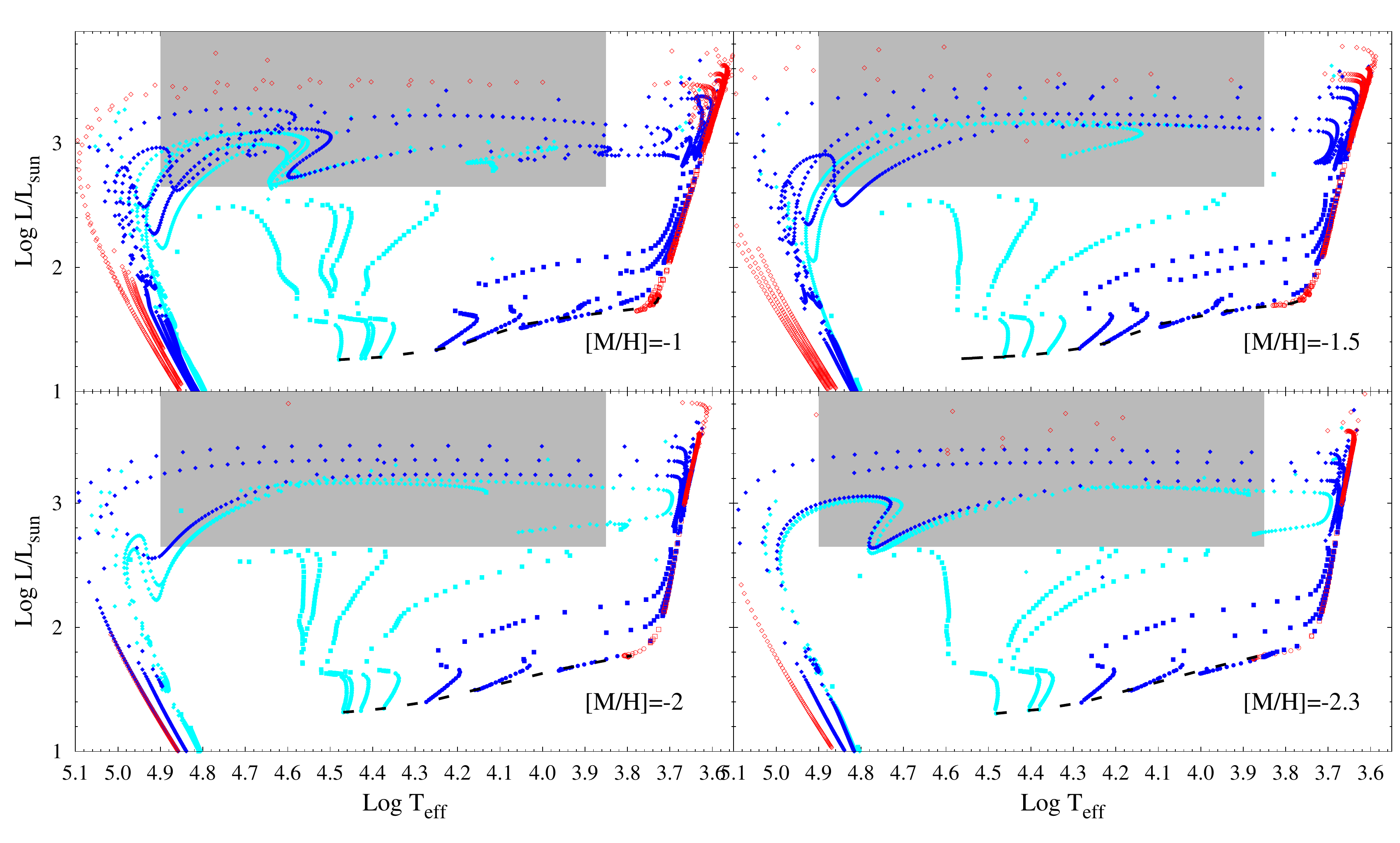}
\caption[]{Stellar evolution sequences computed for the present work
  (see Table\,\ref{tab:sequences} for details). Black circles indicate the
  location of the HB and are plotted with a time step of
  $5$\,Myr, while squares and diamonds indicate the post-HB evolution
  with time steps of 0.5\,Myr and 5\,kyr, respectively. Cyan, blue, and red
  symbols indicate the sequences that populate the extreme, blue, and
  red parts of the HB. The gray region marks the
    parameter space of the hot post(-early) AGB stars in Table\,\ref{tab:result-gc}.}\label{fig:tracks}
\end{center}
\end{figure*}

In order to compare the stellar evolution models with the observed
number of stars, we adopted the evolutionary flux method (see
\citealt{Greggio+2011} for details). This method assumes that the
late evolutionary stages are much shorter than the main-sequence
evolution and that the number of ``dying stars'' can be equated to the
number of stars leaving the main sequence. Under these assumptions, the
number of stars $N_k$ in a simple stellar population at each evolutionary
stage $k$ is given by
\begin{equation}
N_k= B\times L_{\rm total}\times t_k,
\label{eq: evolflux}
\end{equation}
where $t_k$ is the duration of the evolutionary stage $k$,
$L_{\rm total}$ is the total luminosity of the stellar population, and
$B$ stands for the specific evolutionary flux. For old stellar
populations, like the globular clusters studied in this work, we can
approximate $B\simeq 2\times 10^{-11}$ stars per year and per solar
luminosity (see \citealt{Greggio+2011} for details). In our case, we
are interested in the time spent by the models in the region of the
Hertzprung-Rusell ({\bf HR}) diagram where we would observe them as
hot UV-bright post-AGB stars. We define the region in the HR
diagram corresponding to hot UV-bright post-AGB stars detected
in optical colours as that defined by $\log L/L_\odot>2.65$ and
$4.9<\log T_{\rm eff}<3.845$ (grey zone in
Fig.\,\ref{fig:tracks}). Each computed track casts a timescale $t$
corresponding to the time spent in that part of the HR diagram. With
the parameters shown in Table\,\ref{tab:par-gc} and with Eq.\,\ref{eq:
  evolflux} , we can compute for each cluster the expected numbers $N^t$
predicted by each track $t$ of a similar metallicity. 
It should be emphasized that Eq.\,\ref{eq: evolflux} assumes that the
whole cluster population  evolved through that specific
  track $k$. Each post-HB track computed for the cluster metallicity
  gives a different expected number $N_k$ (Eq.\,\ref{eq: evolflux}),
  and the actual number of expected post-HB stars will depend on the
  frequency with which each specific track is followed in each
  cluster. Lacking a better characterization of the HB
  demographics, we can split the HB into the RHB, BHB, and EHB and then
  estimate the number of stars evolving from the RHB, BHB, or EHB
  using the relative populations of the RHB, BHB, and EHB in each
  cluster\footnote{The lifetimes on the HB are not strongly
    affected by the effective temperature of the models, therefore this is a
    decent approximation.} ($f_{\rm RHB}$, $f_{\rm BHB}$, and
  $f_{\rm EHB}$ respectively, see Table\,\ref{tab:par-gc}). Using all
  sequences evolving from the RHB, BHB, or EHB, we obtain a range of
  expected numbers for hot post-AGB stars evolving from each part of
  the HB ($N^{\rm RHB}$, $N^{\rm BHB}$, or $N^{\rm EHB}$, respectively). The actual range of expected UV-bright stars in the
  post-AGB region of the HR diagram (N) of a given cluster is then
  estimated as

\begin{equation}
N= f_{\rm EHB} \cdot N^{\rm EHB} + f_{\rm BHB} \cdot N^{\rm BHB} + f_{\rm RHB}
\cdot N^{\rm RHB}. 
\label{eq: number}
\end{equation}

The resulting ranges of expected numbers for each cluster are shown in
Table\,\ref{tab:result-gc}. We excluded NGC\,2808 from this
  investigation because we know that several luminous hot
  UV-bright stars are not included in our study \citep{Jain+19}.
\begin{figure}
\includegraphics[width=\columnwidth, angle=0]{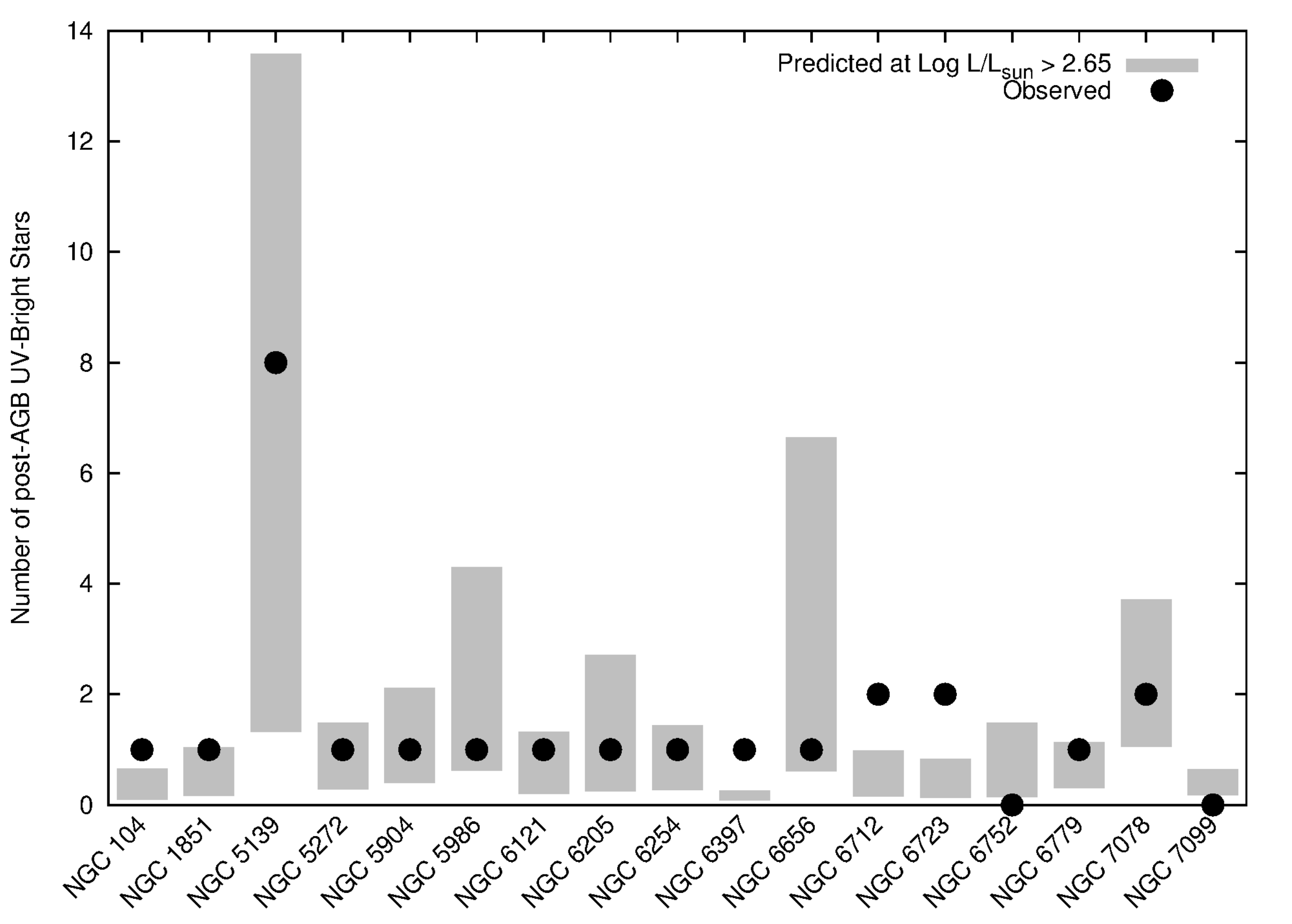}
\caption[]{Numbers of predicted (grey bars) and spectroscopically
    confirmed (black dots) hot post-(early) AGB stars in globular clusters.}
\label{fig:result-gc}
\end{figure}
Overall, the agreement between the number of post-AGB stars estimated
(N$^{\rm post-AGB}_{\rm exp}$, fourth column in
Table\,\ref{tab:result-gc}) and the observed number of post-AGB stars
(N$^{\rm post-AGB}_{\rm obs}$, second column in
Table\,\ref{tab:result-gc}) in each cluster is good (see also
Fig.\,\ref{fig:result-gc}). In 11 out of 17 cases
  ($\approx$65\%) the observed number of stars lies within the
computed range, while in 3 other cases (NGC\,104, NGC\,6752, and
NGC\,7099) the observed number is just an integer number next to the
expected range. A more significant discrepancy is observed for
  NGC\,6712 and NGC\,6723, where the upper boundary of the predicted
  ranges is a factor $\approx$2 lower than the observed number. An
  even larger discrepancy is observed for NGC\,6397, for which the
  largest predicted number is about four times smaller than the
  observed one. In these three cases, the observed value is higher than
  the predicted ones, but a word of caution is in order. We have
  preselected for our sample those clusters that do harbour hot UV-bright
  stars. It is therefore possible that our clusters are biased to
  larger post-AGB numbers. Given the low number of the expected
  post-AGB stars and the fact that the observed number is necessarily an
  integer, it is difficult to point to this discrepancy as a
  discrepancy with the stellar evolution models. A full
  comparison of all clusters searched for hot UV-bright stars is required
  for firmer conclusions.

\section{Conclusions}
\label{sec:conclusions}
Our results first confirm that spectroscopic observations of
UV-bright stars are required to verify their evolutionary status, as is shown by the
cases of NGC\,7099$-$UVBS2 and NGC\,6121$-$UVBS2. The need for
metal-rich model spectra in the analysis of hot evolved stars has been
reported elsewhere \citep{Latour+15} , and we can confirm that
additional opacities are required to reproduce the He\,{\sc i} and
He\,{\sc ii} lines simultaneously and the UV spectra. This
should be kept in mind when the contribution of hot
post-AGB stars to the UV flux of evolved populations is estimated.

We find general agreement for the number of observed hot post-AGB stars compared to
predictions from evolutionary theory,
although the numbers per cluster are low (at most two, except for
$\omega$\,Cen, which has five). Some discrepancies are observed in the
clusters NGC\,6397, NGC\,6712, and NGC\,6723, which show somewhat larger
numbers than expected from the models. Because of the 
  small-number statistics nature of the comparison, it is unclear, however,
whether this is due to a problem with the models or a result of the way the sample
was preselected. In order to improve our constraints on low-mass
post-AGB stellar evolution models, the
number of studied globular clusters needs to be increased. When the full set of 78 globular
clusters with UV-imaging is studied, it will be possible to combine
clusters with similar metallicities and HB morphologies to improve the
number statistics in the comparisons.

\begin{acknowledgements}
  We are grateful to Nicole Reindl, who made her co-added spectrum of
  SDSS\,J155610.40$+$254640.3 and its model spectrum available to us.
  We thank Simon Jeffery for the merger track and his explanations of
  the merger behaviour. We appreciate the anonymous referee's
    comments that improved the readability and clarity of this paper
    substantially.  M.M.M.B. is partially supported through ANPCyT
  grant PICT-2016-0053 and MinCyT-DAAD bilateral cooperation program
  through grant DA/16/07. This research has made use of NASA's
  Astrophysics Data System Bibliographic Services and of the VizieR
  catalogue access tool, CDS, Strasbourg, France. The original
  description of the VizieR service was published in \cite{VizieR}.
  Some of the data presented in this paper were obtained from the
    Mikulski Archive for Space Telescopes (MAST). STScI is operated by
    the Association of Universities for Research in Astronomy, Inc.,
    under NASA contract NAS5-26555.
\end{acknowledgements}
\bibliography{UVBS}
\begin{appendix}
\section{Flux calibration}\label{app:flux}
GRIS\_1200B is a volume-phase
holographic grism. Its response therefore depends on the position of
the slit along the dispersion axis. The best solution for a flux
calibration is to take the flux standard star at the same
place on the detector as the science spectrum. This is done for data taken
with the long slit by creating a 5\arcsec\ wide long slit with the
slit blades of the MOS slitlets. For MOS observations with distributed
slitlets, this would require taking several flux standard star
observations, which is not feasible in service mode.

To correct the response derived from standard stars taken at different
positions than the science targets, we made use of the fact that the
varying response is seen also in the spectral energy distribution of
the screen flat fields taken for the MOS data (see
\citealt{Milvang-Jensen+08} and \citealt{Messenger} for a more detailed
description). Therefore we took the ratio of the wavelength-calibrated
master screen flat and normalized master flat and averaged it along
the slit. Then we divided the extracted standard star spectra and the extracted science spectra by their flat-field
spectra. This procedure will of course introduce the lamp spectrum
into these spectra, but as long as the lamp spectrum does not change,
this can be corrected for with the response curve. 

\begin{figure}
\includegraphics[height=\columnwidth,angle=270]{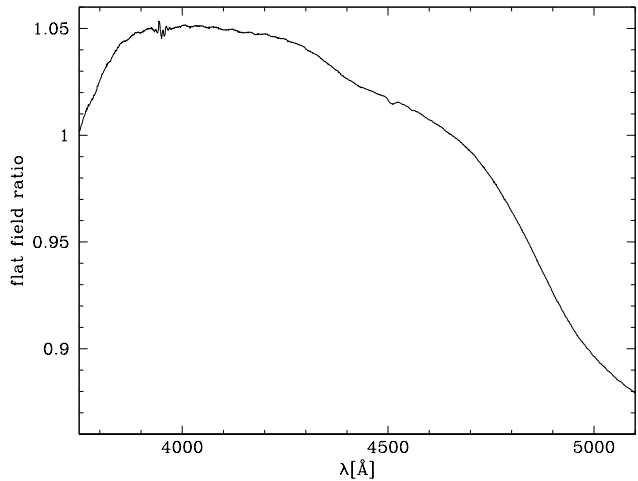}
\caption[]{Ratio of the averaged wavelength-calibrated flat-field spectra
  observed with the LSS 0\farcs5 slit and with a 5\arcsec\ MOS slit at
  the position of the 0\farcs5 long slit (both normalized by their
  median flux). The ratio shows clear large-scale variations of some
  20\%. }\label{fig:specratio}
\end{figure}

During this exercise we noted three potential problems:
\begin{enumerate}
\item The narrow-slit flat fields taken until June 22, 2012, show two
  emission lines at about 3944.2\,\AA\ and 3961.8\,\AA. The lines are
  not visible in the 5\arcsec\ flat fields taken for the standard
  stars. The smoothing of the master screen flat along the dispersion
  axis will smear out these lines to a radius of at least $\pm$10
  pixels or $\pm$7\,\AA\ (relative to the peak position).  We therefore did not use these regions in our later analysis.
\item Between June 22 and July 22, 2012, the spectral energy
  distribution of the flat-field lamps changed.
\item Even between flat fields taken within 30 minutes of each other,
  the spectral energy distribution changed by some 20\% (see
  Fig.\,\ref{fig:specratio}). 
\end{enumerate}

All extracted spectra (science and standard stars) were corrected for
atmospheric extinction using the extinction coefficients of
\citet{Patat+11}.

In order to have a finely sampled response, we used observations of
EG\,274 (April 20, 2012) and Feige\,110 (June 22, 2012), for which
finely sampled reference spectra exist \citep{Moehler+14}. We
determined the radial velocity of the observed spectrum using the
H$\delta$ line, aligned the noise-free reference spectrum to the
same radial velocity, and resampled it to the same wavelength steps as
the observed spectrum. Then we computed the ratio of the observed
extinction-corrected standard star spectrum and the resampled
reference spectrum. This ratio showed small residuals at the positions
of strong stellar lines, which we masked when we fit a smoothed spline
to the ratio. This fit was then used to flux-calibrate the
extinction-corrected science data. 

As noted above, the spectral energy distribution of the flat fields
changed between June 22 and July 22, 2012. Only long-slit spectra were
taken after June 22. We therefore decided to use the flat-field
spectrum taken for the long slit on June 14, 2012, to correct all
long-slit spectra. This procedure allowed us to use the standard star
observed on June 22, 2012 also for data from July. In principle, the
long-slit spectra do not even need to be corrected if a standard star
was observed at the same position (as is the case for
Feige\,110). However, the smoothed flat-field spectra show some
variations in detector response at the blue end that are hard to fit
without a high risk of fitting noise in the response. The division by
the flat-field spectra is therefore useful beyond correcting the
position dependency of the response.

In order to deal with the short-term variations of the flat-field lamp we
used the same flat-field spectrum for the standard star taken at the position
of the LSS 0\farcs5 slit and all science data taken with the long
slit. For the MOS data we could not use the same flat field for
science data and standard star because they were observed at different
positions on the CCD.

\section{Details of the stellar evolution sequences}
\begin{table*}
\caption{Description of the stellar evolution sequences computed for this work. TP-AGB stands for thermally pulsing AGB, (V)LTP for (very) late thermal pulse}\label{tab:sequences}
\begin{tabular}{cccccc}
\hline\hline
$M_{\rm ZAHB}$  & $T_{\rm eff, ZAHB}$ & $\log g_{\rm ZAHB}$ & HB location & post-HB behavior 
& $M_{\rm final}$\\
\hspace*{0mm} [$M_\odot$] & [K] & [cm s$^{-2}$] & & &[$M_\odot$]\\ \hline
\\[-3mm]
\multicolumn{6}{c}{[M/H]$=-1$, Age$= 12$ Gyr, $M_{\rm ZAMS}=0.85 M_\odot$}\\[1mm]
 0.490 & 30138 & 5.74 &  EHB &   post-EHB, no thermal pulses  & 0.490 \\
 0.494 & 26706 & 5.52 &  EHB &   1 thermal pulse (like a LTP) & 0.493\\
 0.495 & 26152 & 5.49 &  EHB &   1 thermal pulse (like a LTP) & 0.495\\
 0.500 & 24040 & 5.34 &  EHB &   2 thermal pulses (like a LTP)& 0.496\\
 0.530 & 17770 & 4.78 &  BHB &   post-EAGB, 2 thermal pulses (like a LTP)& 0.499\\
 0.550 & 15264 & 4.48 &  BHB &   TP-AGB + LTP                 & 0.504\\
 0.580 & 11097 & 3.80 &  BHB &  TP-AGB + LTP    & 0.513\\
 0.600 &  8815 & 3.35 &  BHB &   TP-AGB                       & 0.518\\
 0.650 &  5724 & 2.56 &  RHB &   TP-AGB +LTP                  & 0.528$\ddagger$\\
 0.700 &  5484 & 2.50 &  RHB &   TP-AGB                       & 0.537\\
 0.750 &  5392 & 2.47 &  RHB &   TP-AGB                       & 0.545\\
 0.850 &  5315 & 2.46 &  RHB &   TP-AGB                       & 0.555\\\hline
\\[-3mm]
\multicolumn{6}{c}{ [M/H]$=-1.5$, Age$= 11.7$ Gyr, $M_{\rm ZAMS}=0.83 M_\odot$  }\\[1mm]
 0.495 & 29013 & 5.66 &  EHB &   post-EHB, no thermal pulses  & 0.495\\
 0.500 & 26116 & 5.47 &  EHB &   1 thermal pulse (like a LTP) & 0.499\\
 0.510 & 22942 & 5.23 &  EHB &  post-EAGB, 1 thermal pulse (like a LTP) & 0.500\\
 0.530 & 19297 & 4.91 &  BHB &  post-EAGB, 1 thermal pulse (like a LTP) & 0.501\\
 0.550 & 16897 & 4.66 &  BHB &   TP-AGB + LTP                 & 0.505\\
 0.600 & 12354 & 3.99 &  BHB &   TP-AGB + LTP                 & 0.519\\
 0.650 &  8670 & 3.31 &  BHB &   TP-AGB + VLTP                & 0.525$\ddagger$\\
 0.700 &  6105 & 2.67 &  RR\,Lyr & TP-AGB + LTP               & 0.540$\ddagger$\\
 0.750 &  5700 & 2.54 &  RR\,Lyr & TP-AGB + LTP               & 0.550$\ddagger$\\
 0.830 &  5542 & 2.50 &  RHB & TP-AGB                         & 0.557\\\hline
\\[-3mm]
\multicolumn{6}{c}{ [M/H]$=-2$, Age$= 11.8$ Gyr, $M_{\rm ZAMS}=0.82 M_\odot$  }\\[1mm]
 0.505 & 29455 & 5.65 &  EHB &   post-EHB, no thermal pulses  & 0.505\\
 0.506 & 28773 & 5.61 &  EHB &   post-EHB, no thermal pulses  & 0.506\\
 0.510 & 26760 & 5.48 &  EHB &   1 thermal pulse (like a LTP) & 0.509\\
 0.520 & 23648 & 5.26 &  EHB &   post-EAGB, 1 thermal pulse (like a LTP) & 0.510\\
 0.550 & 18774 & 4.83 &  BHB &   post-EAGB, 1 thermal pulse (like a LTP) & 0.513\\
 0.600 & 14431 & 4.31 &  BHB &   TP-AGB                       & 0.525\\
 0.700 &  9064 & 3.40 &  BHB &   TP-AGB                       & 0.547\\
 0.820 &  6164 & 2.69 &  RR\,Lyr & TP-AGB                     & 0.557\\\hline
\\[-3mm]
\multicolumn{6}{c}{ [M/H]$=-2.3$, Age$= 11.7$ Gyr, $M_{\rm ZAMS}=0.82 M_\odot$  }\\[1mm]
 0.505 & 30346 & 5.72 &  EHB &   post-EHB, no thermal pulses  & 0.505\\
 0.515 & 25384 & 5.39 &  EHB &   1 thermal pulse (like a LTP) & 0.511\\
 0.520 & 24034 & 5.29 &  EHB &  post-EAGB, 1 thermal pulse (like a LTP) & 0.512\\
 0.550 & 19090 & 4.86 &  BHB &   TP-AGB                       & 0.513\\
 0.600 & 14787 & 4.35 &  BHB &   TP-AGB+VLTP                  & 0.510\\
 0.700 &  9681 & 3.51 &  BHB &   TP-AGB                       & 0.546\\
 0.820 &  7967 & 2.92 &  RR\,Lyr & TP-AGB                     & 0.570\\\hline
\multicolumn{6}{c}{\small $\ddagger$ These sequences ended highly H-deficient  due to burning or dilution of the
H-rich envelope during the last He-shell flash.}

\end{tabular}

\end{table*}
\end{appendix}
\end{document}